\documentclass{aa}
\usepackage{psfig}
\usepackage{graphicx}  % standard LaTeX graphics tool
\usepackage{subeqnar}  % subnumbers individual equations
\makeindex             % used for the subject index
\usepackage{feynmp,
amsfonts,amsbsy}

\newcommand{\lsim}{\stackrel{\scriptstyle <}{\phantom{}_{\sim}}}
\newcommand{\gsim}{\stackrel{\scriptstyle >}{\phantom{}_{\sim}}}
\def\lsim{\unitlength5mm\begin{picture}(1,0)
\put(0,.35){\makebox(1,0){$<$}}\put(0,0){\makebox(1,0){$\sim$}}
\end{picture}}
\def\pls{\makebox(0,0){$+$}}

\def\ssp{\makebox(0,0)
    {\thinlines\put(-.1,0){\line(1,0){.2}}\put(0,-.1){\line(0,0){.2}}}}
\def\ssm{\makebox(0,0){\put(-.1,0){\thinlines\line(1,0){.2}}}}
\def\photon{\thinlines\multiput(0,0)(.2,0){3}{\line(1,0){0.1}}}

\def\oneloopvertex{
    \put(1.625,0){\thicklines\oval(2.0,1.5)}
    \put(0,0){\photon}\put(0.35,0.3){\ssp}
    \put(0.625,0){\circle*{.25}}\put(2.625,0){\circle*{.25}}
    \put(2.75,0){\photon}\put(2.9,0.3){\ssm}}

\def\fullbox{\makebox(0,0){\rule{1.5mm}{3mm}}}
\def\interaction{\makebox(0,0){\put(0,0){\interact}
    \put(0,.95){\ssp}\put(0,-.95){\ssm}
    \put(0,.125){\ssp}\put(0,-.125){\ssm}}}
\def\interact{\makebox(0,0){\put(0,.5){\fullbox}
    \thicklines\put(0,0){\oval(.75,.5)}
    \put(0,-.5){\fullbox}} }

\def\til2loop{\put(0,0){\oneloopvertex}\put(4.,0){\pls}
      \put(4.75,0){\oneloopvertex}\put(6.375,0){\interaction}
      \put(9,0){\pls}}

\begin{fmffile}{medneutrd}
\begin{document}

%   \thesaurus{08     % A&A Section 8: Stars
%              (02.04.1;  % Dense matter,
%               08.05.3;  %
%               08.09.3;  % Stars: interiors,
%               08.14.1;} % Stars: neutron.
%
   \title{Medium effects in cooling of neutron stars and the $3P_2$ neutron gap
        %\thanks{Research supported in part by the
%DFG under grant no. 436 RUS 17/117/03}
}

%   \subtitle{}

   \author{ H. Grigorian \inst{1,2}
\and D.N. Voskresensky \inst{3,4}}
%%}

 \offprints{D.N. Voskresensky}

 \institute{Institut f\"ur Physik, Universit\"at Rostock,
        Universit\"atsplatz 3, D--18051 Rostock, Germany\\
        email: hovik.grigorian@uni-rostock.de\\
         \and Department of Physics, Yerevan State University, Alex
        Manoogian Str. 1, 375025 Yerevan, Armenia\\
        \and Gesellschaft f\"ur Schwerionenforschung mbH, Planckstr. 1,
        D--64291 Darmstadt, Germany\\
        email: d.voskresensky@gsi.de\\
        \and Moscow Institute for Physics and Engineering, Kashirskoe sh. 31,
        115409 Moscow, Russia\\
%%email: d.voskresensky@gsi.de
\date{Received: 25. February 2004; accepted: }
}
\abstract{
We study the dependence of the cooling of isolated neutron
stars on the magnitude of the $3P_2$ neutron gap. Our ``nuclear medium
cooling'' 
scenario favors a suppressed value
of the $3P_2$ neutron gap.
    \keywords{Dense baryon matter,
neutron stars,  medium effects,  nucleon gaps, pion softening, heat transport}
}
\maketitle

%
%________________________________________________________________

\section{Introduction} \label{sec:intro}
Theoretical study of neutron star (NS) cooling  began with pioneering
works of
\cite{TC65} and \cite{BW65}. 
It has been argued that  one-nucleon  direct
Urca (DU)
processes $n\rightarrow pe\bar{\nu}$,
$pe\rightarrow n\nu$ are forbidden
up to sufficiently high density and the main
role is played by the two-nucleon modified Urca (MU) processes,
like $nn\rightarrow npe\bar{\nu}$ and $np\rightarrow ppe\bar{\nu}$.
The  ``standard'' scenario of
NS cooling emerged, where the main process
responsible for the cooling  is the modified Urca process MU
$nn\rightarrow npe\bar{\nu}$ calculated using  free one pion
exchange (FOPE) between nucleons, see \cite{FM79}. An order of
magnitude less contribution to the neutrino emissivity is given by
the nucleon bremsstrahlung (NB) processes. The main process among
NB processes responsible for the cooling  is  neutron-neutron
bremsstrahlung $nn\rightarrow nn\nu\bar{\nu}$,  neutron-proton
bremsstrahlung $pn\rightarrow pn\nu\bar{\nu}$ is less effective, and  the
proton-proton one $pp\rightarrow pp\nu\bar{\nu}$ is even less effective. This scenario
explains only the group of slow cooling data. To explain the group
of rapidly cooling data the ``standard'' scenario was supplemented
by  ``exotic'' processes either with a pion
condensate, with a kaon condensate, with hyperons, or
involving DU reactions,
see \cite{T79,ST83} and
references therein. All these processes may occur only for
densities higher than a critical density, $(2\div 6)~n_0$,
depending on the model, where $n_0$ is the nuclear saturation
density.

Pair breaking and formation   (PBF) processes permitted
in nucleon superfluids have been suggested.  \cite{FRS76}
calculated the emissivity of $1S_0$ neutron  pair breaking and
formation (nPBF) process  and \cite{VS87} considered the general
case. Neutron and proton   pair breaking and formation
processes (pPBF) were incorporated within a closed diagram technique
including medium effects.
Numerical estimates are valid both for $1S_0$ and
$3P_2$ superfluids. \cite{SVSWW97} have shown that the inclusion
of  PBF processes into the cooling code may allow one to describe
the ''intermediate cooling'' group of data (even if one
artificially suppresses the effects of the medium). Thus the ''intermediate
cooling'' scenario arose. 

The PBF processes were then incorporated
in the cooling codes of other groups, that worked with the  ``standard
plus exotics'' scenario, see \cite{TTTTT02,YGKLP03,PLPS04}.
Recently \cite{PLPS04} called the approach that incorporates the PFB processes
into the ``standard''
scenario,
the ''minimal
cooling'' paradigm.
Some
papers included the possibility of internal heating that results
in a slowing down of the cooling of old pulsars, see \cite{T04}
and refs. therein. However, because of the simplification
of the consideration, calculations within the
``standard plus exotics'' scenario or within the ''minimal
cooling'' paradigm did not incorporate 
effects of the medium.

The necessity to include in-medium effects in the NS cooling
problem is a rather obvious issue. It is based on requirements 
of condensed matter physics, of the physics of the
atomic nucleus and  heavy ion collisions, see \cite{MSTV90,RW,IKHV01}. The relevance of
in-medium effects for the NS cooling problem has been shown by
\cite{VS84}, (1986), (1987),
\cite{SV87,MSTV90,V00} who calculated emissivity of the MU, NB and PFB
processes taking into account in-medium effects. We  call
these processes medium-modified Urca process (MMU), medium
nucleon bremsstrahlung (MNB) and medium pair breaking and formation  (MPBF) processes. The
efficiency of the developed ``nuclear medium cooling'' scenario
for the description of  NS cooling was demonstrated within the
cooling code  by \cite{SVSWW97} and then by \cite{BGV04}. In the latter
paper it was shown that it is possible to fit the whole set of
cooling data available today. Besides the incorporation of
in-medium effects into the pion propagator and the vertices, it
was also exploited that the $3P_2$ neutron gaps are dramatically
suppressed. The latter assumption
 was motivated by the analysis of the data (see Figs 12, 15, 20 -- 23 of
 \cite{BGV04})
and by recent calculations of the $3P_2$ neutron gaps by
\cite{SF03}. However  more recent work of \cite{KCTZ04} suggested
that the
%%$^3$P$_2$
$3P_2$ neutron pairing gap should be dramatically enhanced, as a
consequence of  the strong softening of the pion propagator.
 Thus the results
of calculations of \cite{SF03} and \cite{KCTZ04}, which both
aim to include medium effects in the evaluation of the
$3P_2$ neutron gaps, are contradictory.

Our aim here is to check the consequences of an enhanced
$3P_2$ neutron pairing gap within the "nuclear medium cooling"
scenario following the work of \cite{BGV04}. The paper is
organized as follows. In section \ref{Nuclear} we start with a brief
recapitulation of the
Landau-Migdal
Fermi liquid approach. Subsection \ref{NN} calculates the $NN$
interaction amplitude. In subsection \ref{virt} we separate the
in-medium pion mode yielding the main contribution to the NN
interaction for $n>n_0$. Subsection \ref{ren} demonstrates the
importance of the  renormalization of the weak interaction in the
medium. In subsection \ref{FOPE} we show internal inconsistencies of the
FOPE model, which is the basis of the "standard" scenario.
 Then in section \ref{med} we recapitulate the
 main processes as they are treated within the 
"nuclear medium cooling" scenario. In section \ref{gaps} we
discuss calculations of the pairing gaps and how gaps affect the
emissivities of different processes. Then in section \ref{reg} we
discuss the cooling model of \cite{BGV04}. In section \ref{Khodel}
we present emissivities of the main processes, the heat capacity
and thermal conductivity contributions in the scenario of
\cite{KCTZ04}, when the neutron processes are assumed to be frozen. In
section \ref{numerical} we show our numerical results. Concluding
remarks are given in section \ref{conclusion}.

\section{Medium effects. Nuclear Fermi liquid} \label{Nuclear}

\subsection{$NN$ interaction. Hard and soft
modes}\label{NN}

At temperatures of interest ($T\ll \varepsilon_{Fn},  \varepsilon_{Fp}$) neutrons
and even protons are only slightly excited above their Fermi seas and all the
processes occur in the close vicinity of the Fermi energies
$\varepsilon_{Fn},  \varepsilon_{Fp}$. Quasiparticle approximation is fulfilled for nucleons.
In such a situation a Fermi liquid approach
seems to be the most efficient one. Within this approach the
long-scale phenomena are treated explicitly whereas short-scale
ones are described by the local quantities expressed via
phenomenological  Landau-Migdal parameters. We deal with the
baryon densities $7~n_0 \gsim n\gsim 0.5~n_0$. The value $n_0 \simeq 0.5 m_{\pi}^3$, where 
$m_{\pi}=140~$MeV is the pion mass, $\hbar=c=1$. 
At such densities related to  internucleon distances $d\sim 1/n^{1/3} \sim
(0.5\div 1.3)~1/m_{\pi}$
processes important for the description of the $NN$ interaction
correspond to typical energies and momenta $\omega , k \lsim$ few
$m_{\pi}$. We call them the long-range processes and treat them explicitly.
These
are nucleon particle-hole processes, $\Delta$ isobar-nucleon hole
processes (since the typical energy of the isobar is of the order of  the 
mass difference $m_{\Delta}-m_N \simeq 2.1~m_{\pi}$)
and processes related to the excitation of the pion  (the typical
excitation energy
is of the order of the
pion mass).  
Thus we explicitly present loop diagrams which depend strongly on the energy and
momentum for the $\omega , k \lsim$ few
$m_{\pi}$ of our interest.

Using the above
argumentation of Fermi liquid theory (see \cite{L56,M67,MSTV90})
the retarded $NN$ interaction amplitude is presented as follows
(see also \cite{V00} for further details)
\begin{eqnarray}\label{NN-ampl}
\setlength{\unitlength}{1mm}
\parbox{10mm}{\begin{fmfgraph}(10,10)
\fmfpen{thick} \fmfleftn{l}{2} \fmfrightn{r}{2}
\fmfpolyn{full}{P}{4} \fmf{fermion}{r1,P1} \fmf{fermion}{P2,r2}
%\fmf{heavy}{P2,r2}
\fmf{fermion}{l2,P3}
%\fmf{heavy}{l2,P3}
\fmf{fermion}{P4,l1}
%\fmfv{l=$j$,l.a=120,l.d=3thick}{P4}
\end{fmfgraph}}
%\,\,\,&=&\,\,\,
 &=& \setlength{\unitlength}{1mm}
\parbox{10mm}{\begin{fmfgraph}(10,10)
\fmfpen{thick} \fmfleftn{l}{2} \fmfrightn{r}{2}
\fmfpolyn{shaded,pull=1.4,smooth}{P}{4} \fmf{fermion}{r1,P1}
\fmf{fermion}{P2,r2}
%\fmf{heavy}{P2,r2}
\fmf{fermion}{l2,P3}
%\fmf{heavy}{l2,P3}
\fmf{fermion}{P4,l1}
\end{fmfgraph}}
%\,\,\,+\,\,\,
+ \setlength{\unitlength}{1mm}
\parbox{25mm}{\begin{fmfgraph}(30,10)
\fmfpen{thick} \fmfleftn{l}{2} \fmfrightn{r}{2}
\fmfpolyn{shaded,pull=1.4,smooth}{P}{4} \fmfpolyn{full}{Pr}{4}
%% legs
\fmf{fermion}{l2,P3}
%\fmf{heavy}{l2,P3}
\fmf{fermion}{P4,l1} \fmf{fermion}{r1,Pr1} \fmf{fermion}{Pr2,r2}
%\fmf{heavy}{Pr2,r2}
%%% internal
\fmf{fermion,left=.5,tension=.5}{Pr4,P1}
\fmf{fermion,left=.5,tension=.5}{P2,Pr3}
%\fmf{heavy,left=.5,tension=.5}{P2,Pr3}
\end{fmfgraph}}
%\,\,\,+\,\,\,
\,\,\,
\nonumber\\
\nonumber\\
&+& \setlength{\unitlength}{1mm}
\parbox{15mm}{\begin{fmfgraph}(35,10)
\fmfpen{thick} \fmfleftn{l}{2} \fmfrightn{r}{2}
\fmfpolyn{shaded,pull=1.4,smooth}{P}{4} \fmfpolyn{full}{Pr}{4}
%% legs
\fmf{fermion}{l2,P3}
%\fmf{heavy}{l2,P3}
\fmf{fermion}{P4,l1} \fmf{fermion}{r1,Pr1} \fmf{fermion}{Pr2,r2}
%\fmf{heavy}{Pr2,r2}
%%% internal
\fmf{fermion,left=.5,tension=.5}{Pr4,P1}
%\fmf{fermion,left=.5,tension=.5}{P2,Pr3}
\fmf{heavy,left=.5,tension=.5}{P2,Pr3}
\end{fmfgraph}}
\end{eqnarray}
where
\begin{eqnarray}\label{irred}
\setlength{\unitlength}{1mm}
\parbox{10mm}{\begin{fmfgraph*}(10,10)
\fmfpen{thick}\fmfleftn{l}{2} \fmfrightn{r}{2}
\fmfpolyn{shaded,pull=1.4,smooth}{P}{4} \fmf{fermion}{r1,P1}
\fmf{fermion}{P2,r2}
%\fmf{heavy}{P2,r2}
\fmf{fermion}{l2,P3}
%\fmf{heavy}{l2,P3}
\fmf{fermion}{P4,l1}
\end{fmfgraph*}}
\,\,\,=\,\,\,
\parbox{10mm}{\begin{fmfgraph*}(10,10)
\fmfpen{thick}\fmfleftn{l}{2} \fmfrightn{r}{2}
\fmfpolyn{hatched,pull=1.4,smooth}{P}{4} \fmf{fermion}{r1,P1}
\fmf{fermion}{P2,r2}
%\fmf{heavy}{P2,r2}
\fmf{fermion}{l2,P3}
%\fmf{heavy}{l2,P3}
\fmf{fermion}{P4,l1}
\end{fmfgraph*}}
\,\,\,+\,\,
\parbox{25mm}{\begin{fmfgraph}(25,10)
\fmfpen{thick}\fmfleftn{l}{2} \fmfrightn{r}{2} \fmfpen{thick}
\fmf{fermion}{l2,ol}
%\fmf{heavy}{l2,ol}
\fmf{fermion}{ol,l1} \fmf{fermion}{r1,o2} \fmf{fermion}{o2,r2}
%\fmf{heavy}{or,r2}
%
%%%\fmfv{d.sh=c,d.filled=full,d.si=2thick}{ol}
%%%\fmfv{d.sh=c,d.filled=full,d.si=2thick}{o2}
\fmf{dbl_wiggly,width=1thin}{ol,o2}
\end{fmfgraph}}\,\,\,\,\,\,.
\end{eqnarray}
The solid line presents the quasiparticle Green function of the  nucleon,
the double-line is that of the
$\Delta$ isobar.\footnote{Diagrams with open particle lines have sense only if
the  quasiparticle approximation is valid. Otherwise one needs to use the closed
  diagram technique developed by \cite{VS87}, \cite{KV95}, (1996).} 
Thus most long-range diagrams of the particle-hole and $\Delta$-nucleon hole
types are explicitly presented in
(\ref{NN-ampl}). Other  long-range terms come from the pion.
The double-wave line in (\ref{irred}) corresponds to the exchange
of the free pion with inclusion of the contributions of the
residual $S$ wave $\pi NN$ interaction and $\pi\pi$ scattering,
i.e. the residual  interaction irreducible to the nucleon
particle-hole and delta-nucleon hole. The latter contributions we have taken into account in
(\ref{NN-ampl}).
 Thus the full particle-hole,
delta-nucleon hole and pion irreducible block (first block in
(\ref{irred})) is by its construction significantly more local than
contributions given by explicitly presented graphs. 
If we decided to calculate the irreducible block in (\ref{irred})
we would need to introduce $\sigma$,
$\omega$, $\rho$ exchanges and to  treat diagrams with multiple pion
lines. All these terms are less energy-momentum dependent for $\omega , k \lsim$ few
$m_{\pi}$ than those we treat explicitly.

Up to now we have not perform
any approximations, except the quasiparticle approximation for the nucleons.
Now we perform the main approximation. 
Using the locality argument, one assumes that the Landau-Migdal parameters, $f_{nn}$,
$f_{np}$ in the scalar channel and $g_{nn}$, $g_{np}$ in  the spin channel, which
parameterize the first block in (\ref{irred}),
are
approximately 
constant. Their  values are extracted  from analysis of experimental
data. In a more extended approach these parameters should certainly be calculated as functions
of the density, neutron and proton concentrations, energy and
momentum.

The part of interaction involving $\Delta$ isobar is  analogously
constructed
\begin{equation} \label{gam1-d}
\setlength{\unitlength}{1mm}\parbox{10mm}{
\begin{fmfgraph}(10,10)
\fmfpen{thick} \fmfset{arrow_len}{2mm} \fmfset{arrow_ang}{30}
\fmfleftn{l}{2} \fmfrightn{r}{2} \fmfforce{(0.7w,0.3h)}{T1}
\fmfforce{(0.7w,0.7h)}{T2} \fmfforce{(0.3w,0.7h)}{T3}
\fmfforce{(0.3w,0.3h)}{T4} \fmfforce{(1.0w,0.0h)}{r1}
\fmfforce{(1.0w,1.0h)}{r2} \fmfforce{(0.0w,1.0h)}{l2}
\fmfforce{(0.0w,0.0h)}{l1} \fmfpolyn{shaded,smooth,pull=1.4}{T}{4}
\fmf{fermion}{r1,T1} \fmf{heavy}{T2,r2} \fmf{fermion}{l2,T3}
\fmf{fermion}{T4,l1}
\end{fmfgraph}}\,\,=\,\,
\setlength{\unitlength}{1mm}\parbox{10mm}{
\begin{fmfgraph*}(10,10)
\fmfpen{thick}\fmfset{arrow_len}{2mm} \fmfset{arrow_ang}{30}
\fmfleftn{l}{2} \fmfrightn{r}{2} \fmfforce{(0.7w,0.3h)}{T1}
\fmfforce{(0.7w,0.7h)}{T2} \fmfforce{(0.3w,0.7h)}{T3}
\fmfforce{(0.3w,0.3h)}{T4} \fmfforce{(1.0w,0.0h)}{r1}
\fmfforce{(1.0w,1.0h)}{r2} \fmfforce{(0.0w,1.0h)}{l2}
\fmfforce{(0.0w,0.0h)}{l1}
\fmfpolyn{hatched,smooth,pull=1.4}{T}{4} \fmf{fermion}{r1,T1}
\fmf{heavy}{T2,r2} \fmf{fermion}{l2,T3} \fmf{fermion}{T4,l1}
\end{fmfgraph*}}
\,\,+\,\, \setlength{\unitlength}{1mm}\parbox{20mm}{
\begin{fmfgraph*}(20,10)
\fmfpen{thick}\fmfset{arrow_len}{2mm} \fmfset{arrow_ang}{30}
\fmfleftn{l}{2} \fmfrightn{r}{2} \fmfforce{(1.0w,0.0h)}{r1}
\fmfforce{(1.0w,1.0h)}{r2} \fmfforce{(0.0w,1.0h)}{l2}
\fmfforce{(0.0w,0.0h)}{l1} \fmf{fermion}{l1,ol}
\fmf{fermion}{ol,l2} \fmf{fermion}{r1,or}\fmf{heavy}{or,r2}
\fmf{dbl_wiggly,width=thin}{ol,or}
%%%\fmfdot{ol,or}
\end{fmfgraph*}}\,\,\,\,\,\,\,\,.
\end{equation}
The main part of the $N\Delta$ interaction is due to the pion
exchange. Although information on the local part of the $N\Delta$
interaction is rather scarce, one can conclude
(\cite{MSTV90,SST99}) that the corresponding Landau-Migdal
parameters are essentially smaller then those for the $NN$
interaction.
Therefore  for simplicity we neglect the first graph in the r.h.s. of
(\ref{gam1-d}).

Straightforward resummation of (\ref{NN-ampl}) in the neutral channel
yields 
(\cite{VS87,MSTV90}) 
\\
\begin{eqnarray}\label{neutr-land}
&&\Gamma_{\alpha\beta}^R =\setlength{\unitlength}{1mm}
\parbox{10mm}{\begin{fmfgraph*}(10,10)
\fmfpen{thick} \fmfleftn{l}{2} \fmfrightn{r}{2}
\fmfpolyn{full}{P}{4} \fmf{fermion}{r1,P1} \fmf{fermion}{P2,r2}
%\fmf{heavy}{P2,r2}
\fmf{fermion}{l2,P3}
%\fmf{heavy}{l2,P3}
\fmf{fermion}{P4,l1}
%\fmfv{l=$j$,l.a=120,l.d=3thick}{P4}
\fmflabel{$\alpha$}{l1}\fmflabel{$\alpha$}{l2}
\fmflabel{$\beta$}{r1}\fmflabel{$\beta$}{r2}
\end{fmfgraph*}}\nonumber\\
\nonumber\\
\nonumber\\
 &&= C_0 \left({\cal{F}}_{\alpha\beta}^R
+{\cal{Z}}_{\alpha\beta}^R {\vec{\sigma}}_1 \cdot{\vec{\sigma}}_2
\right)+f_{\pi N}^2{\cal{T}}_{\alpha\beta}^R (
{\vec{\sigma}}_1\cdot{\vec{k}}) ({\vec{\sigma}}_2\cdot{\vec{k}}),
\end{eqnarray}
%where $\alpha$ and $\beta$ are $n$ or $p$,
\begin{eqnarray}\label{c-f1}
{\cal{F}}_{\alpha\beta}^R &=&f_{\alpha\beta}\gamma
(f_{\alpha\beta}),\,\,\,\, {\cal{Z}}_{nn}^R =g_{nn}\gamma
(g_{nn}),\,\,\,\,\\
 {\cal{Z}}_{np}^R &=&g_{np}\gamma
(g_{nn}),\,\,\,\alpha , \beta =(n,p),
\nonumber \\
{\cal{T}}_{nn}^R &=&\gamma^2 (g_{nn})D^{R}_{\pi^0},\,\,\,\,
{\cal{T}}_{np}^R =-\gamma_{pp}\gamma
(g_{nn})D^{R}_{\pi^0},\,\,\,\,\nonumber\\
{\cal{T}}_{pp}^R &=&\gamma^2_{pp}D^{R}_{\pi^0},\nonumber \\
\gamma^{-1} (x)&=& 1-2 x C_0 A_{nn}^R ,\,\,\,\,\gamma_{pp}=(1-4 g
C_0  A_{nn}^R ) \gamma(g_{nn}),\nonumber
\end{eqnarray}
$f_{nn}=f_{pp}=f+f'$, $f_{np}=f-f'$, $g_{nn}=g_{pp}=g+g'$, and
$g_{np}=g-g'$, the dimensional normalization  factor is usually taken
to be $C_0 =\pi^2 /[m_N p_{FN} (n_0 )]\simeq 300$~MeV$\cdot$ fm$^3$
$\simeq 0.77 m_{\pi}^{-2}$, 
$p_{FN} (n_0 )$ is the nucleon Fermi momentum in
the atomic nucleus, 
$D^{R}_{\pi^0}$ is the full retarded
Green function of $\pi^0$, $A_{\alpha\beta}$ is the corresponding
$NN^{-1}$ loop (without spin degeneracy factor 2)
\\ \\
\begin{eqnarray}\label{loo}
A_{\alpha\beta}&=&\,\,\,
\parbox{10mm}{
\begin{fmfgraph*}(20,10)\setlength{\unitlength}{1mm}
\fmfpen{thick} \fmfleft{l} \fmfright{r}
\fmf{fermion,left=.5,label=$\beta$,label.side=left,left=.8,tension=.8}{l,r}
\fmf{fermion,left=.5,label=$\alpha^{-1}$,
label.side=left,left=.8,tension=.8}{r,l} \,\,
%%%\fmfdot{l}\fmfdot{r}
\end{fmfgraph*}},\\
\nonumber \\
\nonumber\\
 A_{nn}(\omega \simeq q)&\simeq& m^{*2}_n  (4\pi^2
)^{-1} \left( \mbox{ln}\frac{1+v_{Fn}}{1-v_{Fn}} -2v_{Fn}\right)
,\nonumber
\end{eqnarray}
\\
$A_{nn}\simeq -m^*_n p_{Fn}(2\pi^2)^{-1}$, for $\omega \ll
qv_{Fn}, q\ll 2p_{Fn}$, $p_{Fn}=m^*_n v_{Fn}$,  $m^*_n$ is the 
effective neutron mass in medium, and for
simplicity we neglect proton hole contributions due to the low
concentration  of protons. The term proportional to $C_0$ in (\ref{neutr-land})
demonstrates
the change of the local block by the long-range loop $NN$ correlation factors
and the term proportional to the pion-nucleon coupling constant $f_{\pi N}$
demonstrates
the change of the second (pion) term in (\ref{irred}).

Resummation of (\ref{NN-ampl}) in the charged channel yields \\
\begin{eqnarray}\label{ch-land}
\widetilde{\Gamma}_{np}^R &=&\setlength{\unitlength}{1mm}
\parbox{10mm}{\begin{fmfgraph*}(10,10)
\fmfpen{thick} \fmfleftn{l}{2} \fmfrightn{r}{2}
\fmfpolyn{full}{P}{4} \fmf{fermion}{r1,P1} \fmf{fermion}{P2,r2}
%\fmf{heavy}{P2,r2}
\fmf{fermion}{l2,P3}
%\fmf{heavy}{l2,P3}
\fmf{fermion}{P4,l1}
%\fmfv{l=$j$,l.a=120,l.d=3thick}{P4}
\fmflabel{$p$}{l1}\fmflabel{$n$}{l2}
\fmflabel{$p$}{r1}\fmflabel{$n$}{r2}
\end{fmfgraph*}}
= C_0 \left(\widetilde{{\cal{F}}}_{np}^R
+\widetilde{{\cal{Z}}}_{np}^R {\vec{\sigma}}_1
\cdot{\vec{\sigma}}_2 \right) \nonumber\\\nonumber\\
\nonumber\\
 &+&f_{\pi
N}^2\widetilde{{\cal{T}}}_{np}^R ( {\vec{\sigma}}_1\cdot{\vec{k}})
({\vec{\sigma}}_2\cdot{\vec{k}})\,,
\end{eqnarray}
%where
\begin{eqnarray}\label{c-f2}
\widetilde{{\cal{F}}}_{np}^R &=&2f' \widetilde{\gamma} (f'
),\,\,\, \widetilde{{\cal{Z}}}_{np}^R =2g' \widetilde{\gamma} (g'
),\,\,\,\nonumber\\
 \widetilde{{\cal{T}}}_{np}^R
&=&\widetilde{\gamma}^2 (g' )
D^{R}_{\pi^- },\,\,\,\\
\widetilde{\gamma}^{-1} (x)&=& 1-4 x C_0 A_{np}^R \, ,\nonumber
\end{eqnarray}
$D^{R}_{\pi^-}$ is the full retarded
Green function of $\pi^-$.
The Landau-Migdal parameters are poorly known for isospin
asymmetric nuclear matter and for $n >n_0$. Taking into account
the arguments that they can be
considered approximately as constants, for estimates we may use
the values extracted from atomic nucleus experiments.
One  can expect that the most uncertain will be the value of the
scalar constant $f$ due to the essential role of the medium-heavy
$\sigma$ meson in this channel. But this parameter does not enter
the tensor force channel which is the most important in our case.
Unfortunately, there are also  uncertainties in the numerical
values of the  Landau-Migdal parameters in other channels and even for atomic
nuclei. These uncertainties are, mainly, due to attempts to get
the best fit of experimental data in each concrete case, slightly
modifying the parameterization used for the residual part of the $NN$
interaction. For example, calculations by \cite{M67} gave $f\simeq 0.25$,
$f^{\prime} \simeq 1$, $g\simeq 0.5$, $g^{\prime} \simeq 1$
whereas \cite{ST98,FZ95,BTF84}, including quasiparticle
renormalization pre-factors, derived the values $f\simeq 0$,
$f^{\prime} \simeq 0.5 \div 0.6$, $g\simeq 0.05\pm 0.1$,
$g^{\prime} \simeq 1.1 \pm 0.1$.

Typical energies and momenta entering the $NN$ interaction amplitude of 
interest are $\omega\simeq 0$ and $k\simeq p_{Fn}$. Then one
estimates $\gamma (g_{nn} ,\omega\simeq 0 ,k\simeq p_{Fn},
n =n_0 ) \simeq 0.35 \div 0.45$. For $\omega =k \simeq T$ typical
for the weak processes with  $\nu\bar{\nu}$ one
has $\gamma^{-1} (g_{nn} ,\omega\simeq k\simeq T, n =n_0 )\simeq
0.8 \div 0.9$.

\subsection{Virtual pion mode }\label{virt}
The pion is strongly affected by the nucleon surroundings. It is obvious already
from the fact that typical densities are $n\sim 1$ 
and the pion-nucleon coupling constant is $f_{\pi N}=1$ in units
$m_{\pi}=\hbar=c=1$. Obviously perturbation theory is not applicable in this
case. Instead we continue to use the Fermi liquid approach suitable for the description of
the strong interaction.

A straightforward resummation 
of diagrams (\ref{NN-ampl}), (\ref{irred}) yields the following re-summed Dyson equation for
pions
\begin{eqnarray} \label{pion-l}
\parbox{10mm}{\begin{fmfgraph}(20,30)
\fmfleft{l} \fmfright{r}
\fmf{boson,width=thick}{l,r}\fmfforce{(0.0w,0.5h)}{l}
\fmfforce{(1.0w,0.5h)}{r}
\end{fmfgraph}}\,\,&=&\,\,\,\,
\parbox{10mm}{\begin{fmfgraph}(20,30)
\fmfleft{l} \fmfright{r}
\fmf{boson,width=thin}{l,r}\fmfforce{(0.0w,0.5h)}{l}
\fmfforce{(1.0w,0.5h)}{r}\end{fmfgraph}}\,\,+\,\,\,\,
\parbox{25mm}{\begin{fmfgraph}(65,30)
\fmfleft{l} \fmfright{r} \fmf{boson}{l,ol}
\fmf{boson,width=thick}{or,r} \fmfpoly{full}{or,pru,prd}
\fmfforce{(0.2w,0.5h)}{ol} \fmfforce{(0.8w,0.5h)}{or}
\fmfforce{(0.6w,0.3h)}{prd} \fmfforce{(0.6w,0.7h)}{pru}
\fmf{fermion,left=.5,tension=.5,width=1thick}{ol,pru}
\fmf{fermion,left=.5,tension=.5,width=1thick}{prd,ol}\fmfforce{(0.0w,0.5h)}{l}
\fmfforce{(0.9w,0.5h)}{r}
%%%\fmfdot{ol}
\end{fmfgraph}}
\nonumber\\
\nonumber\\
&+&\,\,\,
\parbox{25mm}{\begin{fmfgraph}(60,30)
\fmfleft{l} \fmfright{r} \fmf{boson}{l,ol}
\fmf{boson,width=thick}{or,r}
%\fmfpoly{hatched}{or,pru,prd}
\fmfforce{(0.2w,0.5h)}{ol}
\fmfforce{(0.8w,0.5h)}{or}\fmfforce{(1.0w,0.5h)}{r}\fmfforce{(0.0w,0.5h)}{l}
%\fmfforce{(0.6w,0.3h)}{prd}
%\fmfforce{(0.6w,0.7h)}{pru}
\fmf{heavy,left=.6,tension=.5}{ol,or}
\fmf{fermion,left=.6,tension=.5,width=1thick}{or,ol}
%\fmfdot{or}
\fmfv{decor.shape=circle,decor.filled=full,decor.size=4thick}{or}\fmfforce{(0.0w,0.5h)}{l}
\fmfforce{(1.0w,0.5h)}{r}
\end{fmfgraph}}
 +\,\,
\parbox{40mm}{\begin{fmfgraph*}(60,20)
\fmfleft{l} \fmfright{r} \fmf{boson}{l,ol}
\fmf{boson,width=thick}{or,r} \fmfpoly{empty,label=$\Pi_{\rm
res}^R$}{or,pru,plu,ol} \fmfforce{(0.0w,0.0h)}{l}
\fmfforce{(1.0w,0.0h)}{r} \fmfforce{(0.3w,0.0h)}{ol}
\fmfforce{(0.7w,0.0h)}{or} \fmfforce{(0.7w,0.9h)}{pru}
\fmfforce{(0.3w,0.9h)}{plu}
\end{fmfgraph*}}
\end{eqnarray}
The $\pi N\Delta$ full-dot-vertex includes a phenomenological background correction
due to the presence of the higher resonances, $\Pi_{res}^{R}$ is
the residual retarded pion self-energy that includes the
contribution of all the diagrams which are not presented
explicitly in (\ref{pion-l}), such as $S$ wave $\pi NN$ and $\pi\pi$
scatterings (included by double-wavy line in (\ref{irred})). For zero
temperature the main contribution to $\Pi_{res}^{R}$ is given by the
Weinberg-Tomozawa term which has a simple analytic form. A part of  $\Pi_{res}^{R}$
related to $\pi\pi$ fluctuations is important for the description of the vicinity of the pion
condensation critical point. It is calculated explicitly. Other contributions
are rather small and can be neglected, as follows from the comparison of
the theory predictions with
different atomic nucleus data, see \cite{MSTV90}.

The
full $\pi NN$ vertex takes into account $NN$ correlations
\begin{equation} \label{pion-vert}
\parbox{20mm}{\begin{fmfgraph}(40,30)
\fmfleftn{l}{2} \fmfright{r} \fmfforce{(0.0w,0.0h)}{l1}
\fmfforce{(0.0w,1.0h)}{l2} \fmfforce{(1.0w,0.5h)}{r}
\fmfpoly{full}{pr,pl2,pl1} \fmfforce{(0.7w,0.5h)}{pr}
\fmfforce{(0.3w,0.7h)}{pl2} \fmfforce{(0.3w,0.3h)}{pl1}
\fmf{fermion}{l2,pl2} \fmf{fermion}{pl1,l1} \fmf{boson}{pr,r}
\end{fmfgraph}}\,\,=
\,\,
\parbox{20mm}{\begin{fmfgraph}(40,30)
\fmfleftn{l}{2} \fmfright{r} \fmfforce{(0.0w,0.0h)}{l1}
\fmfforce{(0.0w,1.0h)}{l2} \fmfforce{(1.0w,0.5h)}{r}
\fmf{fermion}{l2,ol,l1} \fmf{boson}{ol,r}
%%%\fmfdot{ol}
\end{fmfgraph}}
\,\,+\,\,
\parbox{28mm}{\begin{fmfgraph}(60,30)
\fmfleftn{l}{2} \fmfright{r} \fmfforce{(0.0w,0.0h)}{l1}
\fmfforce{(0.0w,1.0h)}{l2} \fmfforce{(1.0w,0.5h)}{r}
\fmfpoly{full}{pr,pl2,pl1} \fmfforce{(0.8w,0.5h)}{pr}
\fmfforce{(0.6w,0.7h)}{pl2} \fmfforce{(0.6w,0.3h)}{pl1}
\fmfpen{thick}\fmfpoly{hatched,smooth}{brd,bru,blu,bld}
\fmfforce{(0.2w,0.2h)}{bld} \fmfforce{(0.2w,0.8h)}{blu}
\fmfforce{(0.4w,0.8h)}{bru} \fmfforce{(0.4w,0.2h)}{brd}
\fmf{fermion}{l2,blu} \fmf{fermion}{bld,l1}
\fmf{fermion,left=0.2}{bru,pl2} \fmf{fermion,left=0.2}{pl1,brd}
\fmf{boson,width=1thin}{pr,r}
\end{fmfgraph}}
%%%\,\,+\,\,
%%%\parbox{30mm}{\begin{fmfgraph}(80,30)
%%%\fmfleftn{l}{2}
%%%\fmfright{r}
%%%\fmfforce{(0.0w,0.0h)}{l1}
%%%\fmfforce{(0.0w,1.0h)}{l2}
%%%\fmfforce{(1.0w,0.5h)}{r}
%%%\fmfpoly{full}{pr,pl2,pl1}
%%%\fmfforce{(0.8w,0.5h)}{pr}
%%%\fmfforce{(0.6w,0.7h)}{pl2}
%%%\fmfforce{(0.6w,0.3h)}{pl1}
%%%\fmfpoly{hatched}{brd,bru,blu,bld}
%%%\fmfforce{(0.2w,0.2h)}{bld}
%%%\fmfforce{(0.2w,0.8h)}{blu}
%%%\fmfforce{(0.4w,0.8h)}{bru}
%%%\fmfforce{(0.4w,0.2h)}{brd}
%%%\fmf{fermion}{l2,blu}
%%%\fmf{fermion}{bld,l1}
%%%\fmf{heavy,left=0.2}{bru,pl2}
%%%\fmf{fermion,left=0.2}{pl1,brd}
%%%\fmf{boson}{pr,r}
%%%\end{fmfgraph}}
.
\end{equation}
Therefore the nucleon particle-hole part of $\Pi_{\pi^0}$ is
$\propto \gamma (g_{nn})$ and the nucleon particle-hole part of
$\Pi_{\pi^\pm}$ is $\propto \gamma (g' )$.
The value of the $NN$ interaction in the pion channel is
determined by the full pion propagator at small $\omega$ and $k
\simeq p_{Fn}$, i.e. by the quantity 
\begin{equation}
(\omega^*)^2
(k)=-(D^R_{\pi})^{-1}(\omega =0, k,\mu_\pi ). 
\end{equation}
Typical momenta of
interest  are $ k\simeq p_{Fn}$. Indeed the momenta entering
the $NN$ interaction in MU and MMU processes are $ k= p_{Fn}$, the
momenta governing the MNB are $ k=k_m$ (\cite{VS86}) where the
value $ k=k_m \simeq (0.9\div1)p_{Fn}$ corresponds to the minimum
of $(\omega^*)^2 (k)$. The quantity $\omega^* \equiv \omega^* (k_m
)$ is the {\em{effective pion gap}}, i.e., the
effective pion mass in some sense. The effective pion gap $\omega^*$ demonstrates how much the
virtual (particle-hole) mode with pion quantum numbers is softened
at a given density. The quantity ${\omega^*}^2 (p_{Fn} (n))$ replaces the
value $(m_{\pi}^2 +p_{Fn}^2 )$ in the case of the free pion
propagator used for the calculation of the MU process by
\cite{FM79}. It is different for $\pi^0$ and for $\pi^{\pm}$ since
neutral and charged channels are characterized by different
diagrams permitted by charge conservation, thus also depending on
the value of the pion chemical potential, $\mu_{\pi^{+}}\neq
\mu_{\pi^{-}}\neq 0$, $\mu_{\pi^{0}}=0$. For $T\ll
\varepsilon_{Fn},\varepsilon_{Fp}$, one has $\mu_{\pi^{-}}=\mu_e
=\varepsilon_{Fn}-\varepsilon_{Fp}$, as follows from equilibrium
conditions for the reactions $n\rightarrow p\pi^-$ and $n
\rightarrow pe\bar{\nu}$.

As follows from numerical estimates of different $\gamma$ factors
entering (\ref{neutr-land}) and (\ref{ch-land}), the main
contribution to $NN$ interaction for $n>n_0$ is given by the resummed medium one pion
exchange (MOPE) diagram
\begin{equation}\label{amplMOPE}
\parbox{20mm}{\begin{fmfgraph}(20,20)
\fmfleftn{l}{2} \fmfrightn{r}{2} \fmfpoly{full}{p1,p2,p3,p4}
\fmfforce{(0.0w,0.0h)}{l1} \fmfforce{(0.0w,1.0h)}{l2}
\fmfforce{(1.0w,0.0h)}{r1} \fmfforce{(1.0w,1.0h)}{r2}
%%%%%%
\fmfforce{(0.8w,0.0h)}{p1} \fmfforce{(0.8w,1.0h)}{p2}
\fmfforce{(0.2w,1.0h)}{p3} \fmfforce{(0.2w,0.0h)}{p4}
%%%%%%%
\fmf{fermion}{p4,l1} \fmf{fermion}{l2,p3} \fmf{fermion}{p2,r2}
\fmf{fermion}{r1,p1}
\end{fmfgraph}}\,\,\simeq\,\,\,\,\,\,\,\,\,\,\,\,\,\,
\parbox{20mm}{
\begin{fmfgraph}(20,40)
\fmfleftn{l}{2} \fmfrightn{r}{2} \fmfpoly{full}{ur,ul,uo}
\fmfpoly{full}{dr,do,dl} \fmfforce{(0.0w,0.0h)}{l1}
\fmfforce{(0.0w,1.0h)}{l2} \fmfforce{(1.0w,0.0h)}{r1}
\fmfforce{(1.0w,1.0h)}{r2} \fmfforce{(0.2w,0.8h)}{ul}
\fmfforce{(0.8w,0.8h)}{ur} \fmfforce{(0.5w,0.7h)}{uo}
\fmfforce{(0.2w,0.2h)}{dl} \fmfforce{(0.8w,0.2h)}{dr}
\fmfforce{(0.5w,0.3h)}{do} \fmf{fermion}{l2,ul}
\fmf{fermion}{ur,r2} \fmf{fermion}{dl,l1} \fmf{fermion}{r1,dr}
\fmf{boson,width=1thick}{do,uo}
\end{fmfgraph}}\,\,
\end{equation}
if this channel (${\cal{T}}\propto  (
{\vec{\sigma}}_1\cdot{\vec{k}}) ({\vec{\sigma}}_2\cdot{\vec{k}})$, see
(\ref{ch-land}), (\ref{neutr-land}))
of the reaction is not forbidden or suppressed by  specific
effects such as  symmetry, small momentum transfer, etc.

The density dependence of the effective pion gap $\omega^*$ that
we use in this work is demonstrated in Fig. 1 (Fig. 1 of
\cite{BGV04}).
%%%%%%%%%%%%%%%%%  Figure 1 %%%%%%%%%%%%%%%
\begin{figure}[htb]
\psfig{figure=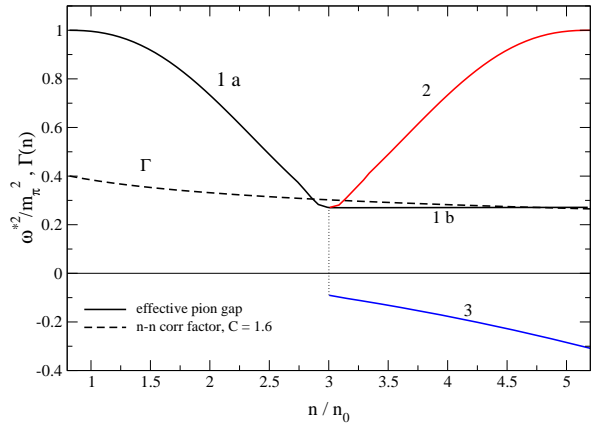,width=0.5\textwidth,angle=-90}
\caption{Nucleon - nucleon correlation factor $\Gamma$ and square
of the effective pion gap $\omega^*$ with pion condensation
(branches 1a, 2, 3) and without (1a, 1b).
%[omegatil.ps]
\label{fig1} }
\end{figure}
%%%%%%%%%%%%%%%%%%%%%%%%%%%%%%%%%%%%%%%%%%
The curve 1a in Fig.1 shows the behavior of the pion gap for
$n<n_c^{\rm PU}$, where $n_c^{\rm PU}$ is the critical density for
the pion condensation. In this work for simplicity  we do not
distinguish between different possibilities of 
$\pi^0$, $\pi^{\pm}$ condensations  and the so-called alternative-layer-structure,
 see  
\cite{VS84,MSTV90} and \cite{UNTMT94}. Thus we assume that $\pi^0$ and $\pi^-$
condensations occur at the very same critical density $n_c^{\rm PU}$.
Although the
value $n_c^{\rm PU}$ depends on  rather uncertain different
parameters we, following \cite{BGV04},  further assume $n_c^{\rm
PU}\simeq 3~n_0$, cf. discussion by \cite{MSTV90}. The curve 1b
demonstrates the possibility  of a saturation of pion softening
and the absence of pion condensation for $n>n_c^{\rm PU}$ (this
possibility could be realized, e.g., if Landau-Migdal parameters
increased with the density). This pion gap (from curves 1a$+$1b) is the value that determines the
pion Green function for the
pion excitations.
Curves 2, 3 demonstrate the
possibility of pion condensation for $n>n_c^{\rm PU}$. The
continuation of  branch 1a  for $n>n_c^{\rm PU}$, called  branch 2,
shows the reconstruction of the pion dispersion relation in the presence of the
condensate state.  
This pion gap 
is the value that determines the
pion Green function for the
pion excitations above the pion condensate vacuum. 
In the presence of the pion condensate (for $n>n_{c\pi}$) the value $\omega^*$
from curve 2 enters the emissivities of all processes with pion excitations in
initial,
intermediate and final reaction states. 

In agreement with the general
trend known in condensed matter physics, fluctuations dominate in
the vicinity of the critical point of the phase transition and die
out far away from it.
The jump from branch 1a to 3 at $n=n_c^{\rm PU}$ is due to the first order phase
transition to the $\pi$ condensation, see discussion of this point
by \cite{D75}, (1982), \cite{VM81}, (1982), \cite{MSTV90}. The $|\omega^*|$
value on 
branch 3  is proportional to the amplitude of the
pion condensate mean field (the line with the cross in the standard notation of
the diagram technique). 

The observation
that pion condensation appears by the first order phase
transition needs a comment. The first order phase transitions
in the systems with several charged species is associated with the
possibility of a mixed phase, see \cite{G92}. The emissivity is
increased within the mixed phase since  efficient  DU-like
processes due to nucleon re-scattering on the new-phase droplets
are possible. However \cite{VYT02}, (2003) and \cite{MTVTC03}, (2005a)
demonstrated that, if it exists, the mixed phase is probably realized
only in a narrow density interval due to  charge screening
effects. Thereby to simplify the consideration we further
disregard the possibility of a mixed phase. We also disregard
the change in the equation of state (EoS) due to  pion condensation assuming that the phase
transition is rather weak.

The density dependence of the correlation factor $\Gamma$ is
presented in Fig. 1. For the correlation factor entering the
emissivity of the MMU process one can use an approximate expression
 $\Gamma (n)\simeq 1/[1+C
(n/n_0 )^{1/3}]$, $C\simeq 1.4\div 1.6$. Note that this  value
$\Gamma$ is an averaged quantity ($\Gamma^6=\Gamma^2_{
w-s}\Gamma^4_s$).  Actually the correlation factor $\Gamma^6$
 entering the emissivity
of the MMU process (see eq. (\ref{MMU}) below) 
looks more involved and depends on the energy-momentum transfer, being
different for vertices connected to the weak coupling (the
correlation factor related to the weak coupling vertices
$\Gamma_{w-s}$ is close to unity) and for vertices related
to the pure strong coupling ($\Gamma_{s}$ is slightly less than
above introduced factor $\Gamma$), see estimates of the energy-momentum dependence
of the correlation factors at the end of the previous subsection.
We see that vertices are rather strongly suppressed (and this
suppression increases with the density) but the softening of the
pion mode is enhanced (${\omega^*}^2 < m_{\pi}^2$) for $n>n_{c1}\simeq
0.5\div 0.8 ~n_0$. Such a behavior  is supported both
theoretically and by analysis of nuclear experiments, see
\cite{MSTV90}. 

Even with full microscopic calculations the
functions $\Gamma (n)$ and $\omega^* (n)$ contain large
uncertainties. These uncertainties come mainly from
simplifications inherent to the Landau-Migdal approach to nuclear
forces where the Landau-Migdal parameters are constants.
However it seems misleading to disregard medium effects only
based on existing rather large uncertainties. 
We believe that further more detailed comparison of theory and experiment will
allow one to reduce these uncertainties.

\subsection{Renormalization of the weak interaction.}\label{ren}

The full weak coupling vertex that takes into account $NN$
correlations is determined by  (\ref{pion-vert}) where now the
wavy line should be replaced by the lepton pair. Thus for the
vertex of interest, $N_1 \rightarrow N_2 l \bar{\nu}$, we
obtain, see \cite{VS87,MSTV90},
\begin{eqnarray}\label{betav}
V_\beta =\frac{G}{\sqrt{2}}\left[ \widetilde{\gamma}(f' )l_0 -g_A
\widetilde{\gamma}(g' ) \vec{l}\vec{\sigma} \right],
\end{eqnarray}
for the $\beta$ decay and
\begin{eqnarray}\label{nn-cor}
V_{nn} &=&-\frac{G}{2\sqrt{2}}\left[ \gamma (f_{nn} )l_0
-g_A\gamma (g_{nn} ) \vec{l}\vec{\sigma} \right],\,\,\,
\nonumber\\
V_{pp}^N &=&\frac{G}{2\sqrt{2}}\left[ \kappa_{pp} l_0
-g_A\gamma_{pp} \vec{l}\vec{\sigma} \right],
\end{eqnarray}
\begin{eqnarray}\label{pp-cor}
\kappa_{pp}&=&c_V -2f_{np}\gamma (f_{nn})C_0
A_{nn},\,\,\nonumber\\
\gamma_{pp}&=&\left( 1 -4g C_0 A_{nn} \right) \gamma (g_{nn}),
\end{eqnarray}
for processes on the neutral currents $N_1 N_2 \rightarrow N_1 N_2
\nu \bar{\nu}$, $V_{pp}=V_{pp}^N +V_{pp}^{\gamma}$, $G\simeq
1.17\cdot 10^{-5}$~GeV$^{-2}$ is the Fermi weak coupling constant,
$c_V =1-4\sin^2 \theta_W$, $\sin^2 \theta_W \simeq 0.23$,
$g_A\simeq 1.26$ is the axial-vector coupling constant, and $l_\mu
=\bar{u}(q_1 )\gamma_\mu (1-\gamma_5 )u(q_2 )$ is the lepton
current. The pion contribution $\sim \vec{q}^{\,2}$ is small for
typical $\mid\vec{q}\mid \simeq T$ or $p_{Fe}$, and for simplicity
is omitted. In medium the value of $g_A$ (i.e. $g_A^*$) slightly
decreases with the density.

The $\gamma$ factors renormalize the corresponding vacuum
vertices. These factors are different for different
processes. The matrix elements of the
neutrino/antineutrino scattering processes $N\nu\rightarrow N\nu$
and of MNB behave differently depending on the energy-momentum
transfer and whether $N=n$ or $N=p$ in the weak coupling vertex.
Vertices\\
\\
\begin{eqnarray}\label{nu-scat}
\,\,\,\,\,\,\,\,\,\,\,\,\,\,\,\,\,\,\,\,\,\,\,\,
\parbox{10mm}{\begin{fmfgraph*}(50,35)
\fmfpen{thick} \fmfleft{l1,l2} \fmfright{r1,r2}
\fmf{fermion}{l1,T2} \fmf{fermion}{T3,r1}
\fmf{fermion,width=1thin}{l2,o3,r2}
\fmf{dashes,width=1thin}{T1,o3}
%%%\fmfpoly{shaded,width=1thin}{p1,p2,p3,p4}
%%%\fmfblob{.05w}{o1}
\fmfpoly{full}{T2,T3,T1} \fmflabel{$\nu$}{l2}
\fmflabel{$\nu$}{r2}\fmflabel{$N$}{l1}\fmflabel{$N$}{r1}\end{fmfgraph*}}
\,\,\,\,\,\,\,\,\,\,\,\,\,\,\,\,\,\,\,\,\,\,\,\, ,
\,\,\,\,\,\,\,\,\,\,\,\,\,\,\,\,\,\,\,\,\,\,\,\,
\parbox{10mm}{\begin{fmfgraph*}(50,35)
\fmfpen{thick} \fmfleft{l1} \fmfright{r1,r2,r3}
\fmf{fermion}{l1,T2} \fmf{fermion}{T3,r1}
\fmf{fermion,width=1thin}{o3,r3}\fmf{fermion,width=1thin}{r2,o3}
\fmf{dashes,width=1thin}{T1,o3}
%%%\fmfpoly{shaded,width=1thin}{p1,p2,p3,p4}
%%%\fmfblob{.05w}{o1}
\fmfpoly{full}{T2,T3,T1} \fmflabel{$\nu$}{r3}
\fmflabel{$\bar{\nu}$}{r2}\fmflabel{$N$}{l1}\fmflabel{$N$}{r1}\end{fmfgraph*}}
\,\,\,
\end{eqnarray}\\
\\
are modified by the correlation factors (\ref{c-f1}) and
(\ref{c-f2}). For $N=n$ these are $\gamma (g_{nn}, \omega ,q)$ and
$\gamma (f_{nn}, \omega ,q)$ leading to an enhancement  of the
cross sections for $\omega > qv_{Fn}$ and to a suppression for
$\omega < qv_{Fn}$. 

Renormalization of the proton vertex (vector
part of $V_{pp}^N +V_{pp}^{\gamma}$) is governed by the processes,
see \cite{VS87,VKK98},\\
\begin{eqnarray}\label{p-vert}
\,\,\,\,\,\,\,\,\,\,\,\,\,\,\,\,
&&\parbox{30mm}{\begin{fmfgraph*}(100,60) \fmfpen{thick}
\fmfleft{l1} \fmfright{r1,r2,r3}
\fmf{fermion,label=$p$,label.side=right}{l1,T2}
\fmf{fermion,label=$p$,label.side=right}{T3,r1}
\fmf{fermion,width=1thin}{o3,r3}\fmf{fermion,width=1thin}{r2,o3}
%\fmf{fermion,width=1thin}{r2,o3}\fmf{fermion,width=1thin}{o3,r3}
\fmf{fermion,label=$n$,label.side=left,left=.8,tension=.8}{T1,o2}
\fmf{fermion,label=$n^{-1}$,label.side=left,left=.8,tension=.8}{o2,T4}
%\fmf{fermion}{o1,o2,o1}
\fmf{dashes,width=1thin}{o2,o3}
%%%\fmfpoly{shaded,width=1thin}{p1,p2,p3,p4}
%%%\fmfblob{.05w}{o1}
\fmfpolyn{hatched,smooth,pull=1.4}{T}{4}
%\fmfdot{o2}
%%\fmflabel{$p$}{l1}
%%\fmflabel{$p$}{r1}
\fmflabel{$\nu$}{r3} \fmflabel{$\bar{\nu}$}{r2}\end{fmfgraph*}}
\,\,\,\,\,\,\,\,\,\,\,\,\,\,\,\,\,\,\,\, \nonumber\\
\nonumber\\
 &+& \,\,\,\,\,\,\,\,\,
\parbox{30mm}{\begin{fmfgraph*}(100,80)
\fmfpen{thick} \fmfleft{l1} \fmfright{r1,r2,r3}
\fmf{fermion,label=$p$,label.side=right}{l1,T2}
\fmf{fermion,label=$p$,label.side=right}{T2,r1}
\fmf{fermion,width=1thin}{o3,r3}\fmf{fermion,width=1thin}{r2,o3}
%\fmf{fermion,width=1thin}{r2,o3}\fmf{fermion,width=1thin}{o3,r3}
\fmf{fermion,width=1thin,label=$e$,label.side=left,left=.8,tension=.8}{T1,o2}
\fmf{fermion,width=1thin,
label=$e^{-1}$,label.side=left,left=.8,tension=.8}{o2,T1}
%\fmf{fermion}{o1,o2,o1}
\fmf{dashes,width=1thin}{o2,o3}\fmf{photon,width=1thick,
label=$\gamma_m$,label.side=left}{T2,T1} \fmflabel{$\nu$}{r3}
\fmflabel{$\bar{\nu}$}{r2}\end{fmfgraph*}}\,\,\,\,\,\,\,\,\,
\,\,\,\,\,\,\,\,\,+\,...
\end{eqnarray}
being forbidden in vacuum but permitted in medium. For the systems with $1S_0$
proton--proton pairing, $\propto g_A^2$ contribution to the
squared matrix element (see (\ref{nn-cor})) is compensated by the
corresponding contribution of the diagram with anomalous Green
functions of protons. The vector current term is $\propto c_V^2$
in vacuum whereas it is $\propto \kappa_{pp}^2$ in medium (by the
first diagram in (\ref{p-vert})). Thereby the corresponding vertices
with protons are enhanced  in medium compared to
small vacuum values ($\propto c_V^2 \simeq 0.006$) leading to
enhancement of the cross sections, up to $\sim 10\div10^2$ times
for $1.5\div3n_0$ depending on the choice of the parameters. It does not
contradict  the statement that correlations
are suppressed in the weak interaction vertices
at $n \leq n_0$. Enhancement with  density comes from
estimation (\ref{loo}) (that directly follows from equation (2.30)
of \cite{VS87}). Also as enhancement factor (up to $\sim 10^2$) comes
from the virtual in-medium photon ($\gamma_m$), second diagram in (\ref{p-vert}), whose propagator
contains $1/m_{\gamma}^2 \sim 1/e^2$, where $m_{\gamma}$ is the
effective spectrum gap, that compensates for the small $e^2$ factor from
electromagnetic vertices, see \cite{VKK98,L00}. It is included by
replacing $V_{pp}^N \rightarrow V_{pp}$. Other processes
permitted in intermediate states like processes with $pp^{-1}$ and
with pions are suppressed by a low proton density and by
$q^2\sim T^2$ pre-factors, respectively. The first diagram in
(\ref{p-vert}) was considered in \cite{VS87}, where the MpPBF
process was first introduced, and then in \cite{MSTV90,SVSWW97}, and it
was shown that MnPBF and MpPBF processes may give contributions of
the same order of magnitude. Finally, with electron-electron hole  (second diagram
(\ref{p-vert})
and neutron-neutron hole   (first diagram (\ref{p-vert}))
correlations included we recover this statement and the numerical
estimate of \cite{VS87,SVSWW97}.

Work of \cite{KV99} gives another example  demonstrating that,
although the vacuum branching ratio of the kaon decays is
$\Gamma(K^-\rightarrow e^-+\nu_e)/\Gamma(K^-\rightarrow
\mu^-+\nu_\mu)\approx 2.5\times 10^{-5}$, in medium (due to
the lambda-proton hole, $\Lambda p^{-1}$, decays of virtual $K^-$) it becomes of
the order of unity. Thus we again see that depending what
reaction channel is considered, in-medium effects may significantly
enhance the reaction rates or substantially suppress them.
Ignoring  these effects may lead to quite misleading results.

\subsection{Inconsistencies of the FOPE model}\label{FOPE}

Since the FOPE model is the basis of the "standard" scenario for
cooling simulations, which is still in use, we would like  to
demonstrate inconsistencies in the model for the
description of interactions in dense ($n \gsim n_0$) baryon medium
(see \cite{VS86,V00}).

The only diagram in the FOPE model which contributes to the MU and NB
is\\
%to the neutrino emissivity of two-nucleon processes is as follows
\begin{eqnarray}\label{FOPEdiag}
\parbox{20mm}{\begin{fmfgraph*}(100,50)
%\fmfpen{thick}
\fmfleft{l1,l2} \fmfright{r1,r2,r3,r4} \fmf{fermion}{l1,o1,r1}
\fmf{fermion}{l2,o2}\fmf{fermion,label=$f_{\pi
N}$,label.side=left}{o2,o3} \fmf{fermion}{o3,r2}
\fmf{fermion}{r3,o4}\fmf{fermion}{o4,r4}
\fmf{dots,width=1thin}{o1,o2}\fmf{dashes,width=1thin}{o3,o4}
%\fmfdot{o1,o2}
%%\fmflabel{$f_{\pi N}$}{o2}
\fmflabel{$f_{\pi N}$}{o1}
\end{fmfgraph*}}
\end{eqnarray}\\
Dots symbolize FOPE. This is the first available Born approximation
diagram, i.e. the second order perturbative contribution in $f_{\pi
N}$ coupling. In order to be theoretically consistent one should use
perturbation theory up to  the same second order in $f_{\pi
N}$ for all the quantities. The pion spectrum is then determined
by a pion polarization operator expanded up to the same order
in $f_{\pi N}$:
\begin{eqnarray}\label{pio}
&&\omega^2 \simeq m_{\pi}^2 +k^2 +\Pi^R_0 (\omega , k, n ),
\,\,\,\,\, \\
\nonumber\\
&&\Pi^R_0 (\omega , k, n )= \,\,\,\,\,\,\,\,\,\,
\parbox{50mm}{
\begin{fmfgraph*}(50,20)
%\fmfpen{thick}
\fmfleft{l} \fmfright{r} \fmf{fermion,left=.8,tension=.8}{o1,o2}
\fmf{fermion,left=.8,tension=.8}{o2,o1}\fmf{dots, label=$f_{\pi
N}$,label.side=left}{l,o1} \fmf{dots,label=$f_{\pi
N}$,label.side=right}{o2,r} \fmfforce{(0.0w,0.5h)}{l}
\fmfforce{(0.2w,0.5h)}{o1} \fmfforce{(0.8w,0.5h)}{o2}
\fmfforce{(1w,0.5h)}{r}\nonumber
%\fmflabel{$f_{\pi N}$}{o2}\fmflabel{$f_{\pi N}$}{o1}
%%%\fmfdot{l}\fmfdot{r}
\end{fmfgraph*}}
\end{eqnarray}\\
The value $\Pi^0 (\omega , k, n )$ is easily calculated containing
no uncertain parameters. For $\omega \rightarrow 0$, $k\simeq
p_F$ and for isospin symmetric matter
\begin{eqnarray}
&&\Pi^R_0
%(\omega \rightarrow 0, k\simeq p_F )
\simeq -\alpha_0 -i \beta_0 \omega ,\,\, \\
&&\alpha_0 \simeq \frac{2m_N p_F k^2 f_{\pi N}^2} {\pi^2}>0,
\,\,\beta_0 \simeq\frac{m_N^2 k f_{\pi N}^2}{\pi}>0.\nonumber
\end{eqnarray}
Replacing this value in  (\ref{pio}) we obtain a solution
$i\beta_0 \omega \simeq (\omega^* )^2 (k)$ with $\mbox{Im} \,\,\omega
<0$ (for $k\simeq k_m$) already for $n>0.3 n_0$ that would mean
appearance of pion condensation ($(\omega^* )^2 (k_m)<0$).
Indeed, the mean field begins to grow as
$\varphi \sim \mbox{exp}({-\mbox{Im} \,\omega \cdot t}) \sim
\mbox{exp}({\alpha t})$, $\alpha >0$, until repulsive $\pi\pi$
interaction does not stop its growth. But it is experimentally
proven that there is no pion condensation in atomic nuclei, i.e.
even at  $n =n_0$. 

The puzzle is solved as follows. The FOPE model
does not work for such densities. One should replace FOPE by the
full $NN$ interaction given by (\ref{neutr-land}),
(\ref{ch-land}). The essential part of this interaction is due to MOPE
with vertices corrected by $NN$ correlations, see
(\ref{amplMOPE}). Also the particle-hole, $NN^{-1}$, part of the pion polarization
operator is corrected by $NN$ correlations. Thus
\begin{equation}
\setlength{\unitlength}{1mm}\parbox{20mm}{\begin{fmfgraph}(20,10)
\fmfleft{l} \fmfright{r} \fmf{boson}{l,ol}
\fmf{boson,width=thin}{or,r} \fmfpoly{full}{or,pru,prd}
\fmfforce{(0.2w,0.5h)}{ol} \fmfforce{(0.8w,0.5h)}{or}
\fmfforce{(0.6w,0.3h)}{prd} \fmfforce{(0.6w,0.7h)}{pru}
\fmf{fermion,left=.5,tension=.5,width=1thick}{ol,pru}
\fmf{fermion,left=.5,tension=.5,width=1thick}{prd,ol}
%%%\fmfdot{ol}
\end{fmfgraph}}\,\,\,\simeq \,\,\, \Pi^R_0 (\omega , k, n )\gamma (g' ,
\omega , k, n )
\end{equation}
is suppressed by the factor $\gamma (g' ,\omega =0, k\simeq p_F
, n \simeq n_0 )\simeq 0.35 \div 0.45$. The final solution of the
dispersion relation (\ref{pio}), now with full $\Pi$ instead of
$\Pi^0$, yields $\mbox{Im}\,\omega >0$ for $n =n_0$ whereas the
solution with  $\mbox{Im} \,\omega <0$, which shows the beginning
of pion condensation, appears only for $n > n_{c\pi}> n_0$.

\section{Medium effects in neutrino radiation processes}\label{med}

\subsection{Medium effects in two-nucleon
processes}
Medium effects essentially modify the contributions of
all processes. The main contributing diagrams are
\\
\begin{eqnarray}\label{MMU-diag}
&&\parbox{30mm}{
\begin{fmfgraph*}(60,80)
\fmfleftn{l}{2} \fmfrightn{r}{4} \fmfpoly{full}{ur,ul,uo}
\fmfpoly{full}{dr,do,dl}\fmfpen{thick} \fmfforce{(0.0w,0.0h)}{l1}
\fmfforce{(0.0w,1.0h)}{l2} \fmfforce{(1.0w,0.0h)}{r1}
\fmfforce{(1.0w,1.0h)}{r2}\fmfforce{(1.2w,0.3h)}{r3}\fmfforce{(1.2w,0.7h)}{r4}
\fmfforce{(0.8w,0.5h)}{o2}\fmfforce{(.5w,0.5h)}{o1}
\fmfforce{(0.4w,0.8h)}{ul} \fmfforce{(0.6w,0.8h)}{ur}
\fmfforce{(0.5w,0.7h)}{uo} \fmfforce{(0.4w,0.2h)}{dl}
\fmfforce{(0.6w,0.2h)}{dr} \fmfforce{(0.5w,0.3h)}{do}
\fmf{fermion}{l2,ul} \fmf{fermion}{ur,r2} \fmf{fermion}{l1,dl}
\fmf{fermion}{dr,r1} \fmf{boson,width=1thick,label=$\pi^0$}{do,o1}
\fmf{boson,width=1thick,label=$\pi^+$}{o1,uo}
\fmf{fermion,width=1thin}{r3,o2}
\fmf{fermion,width=1thin}{o2,r4}\fmf{dashes,width=1thin}{o1,o2}
\fmflabel{$\bar{\nu}$}{r3}\fmflabel{$e$}{r4}\fmflabel{$p$}{r2}
\fmflabel{$n$}{r1}\fmflabel{$n$}{l1}\fmflabel{$n$}{l2}
\end{fmfgraph*}}\,+\,\,\,\,\,\,\,\,\,
\parbox{30mm}{
\begin{fmfgraph*}(60,80)
\fmfleftn{l}{2} \fmfrightn{r}{4} \fmfpoly{full}{ur,ul,uo}
\fmfpoly{full}{dr,do,dl}\fmfpen{thick} \fmfforce{(0.0w,0.0h)}{l1}
\fmfforce{(0.0w,1.0h)}{l2} \fmfforce{(1.0w,0.0h)}{r1}
\fmfforce{(1.0w,1.0h)}{r2}\fmfforce{(1.2w,0.3h)}{r3}\fmfforce{(1.2w,0.7h)}{r4}
\fmfforce{(0.9w,0.5h)}{o2}\fmfforce{(0.65w,0.5h)}{o1}
\fmfforce{(0.4w,0.9h)}{ul}
\fmfforce{(0.6w,0.9h)}{ur}\fmfforce{(0.5w,0.8h)}{uo}\fmfforce{(0.5w,0.7h)}{o4}
\fmfforce{(0.4w,0.1h)}{dl} \fmfforce{(0.6w,0.1h)}{dr}
\fmfforce{(0.5w,0.2h)}{do}\fmfforce{(0.5w,0.3h)}{o3}
\fmf{fermion}{l2,ul} \fmf{fermion}{ur,r2} \fmf{fermion}{l1,dl}
\fmf{fermion}{dr,r1} \fmf{boson,width=1thick,label=$\pi^0$}{do,o3}
\fmf{boson,width=1thick,label=$\pi^+$}{o4,uo}
\fmf{fermion,width=1thin}{r3,o2}
\fmf{fermion,width=1thin}{o2,r4}\fmf{dashes,width=1thin}{o1,o2}
\fmflabel{$\bar{\nu}$}{r3}\fmflabel{$e$}{r4}\fmflabel{$p$}{r2}
\fmflabel{$n$}{r1}\fmflabel{$n$}{l1}\fmflabel{$n$}{l2}
\fmfpoly{full}{o1,o5,o6}
\fmfforce{(0.6w,0.6h)}{o5}\fmfforce{(0.6w,0.4h)}{o6}
\fmfpoly{full}{or,pru,prd} \fmfforce{(0.5w,0.3h)}{or}
\fmfforce{(0.6w,0.4h)}{prd} \fmfforce{(0.4w,0.4h)}{pru}
\fmfpoly{full}{o4,pru1,prd1} \fmfforce{(0.5w,0.7h)}{o4}
\fmfforce{(0.6w,0.6h)}{prd1} \fmfforce{(0.4w,0.6h)}{pru1}
\fmf{fermion,left=.5,tension=.5,width=1thick} {prd1,prd}
\fmf{fermion,left=.5,tension=.5,width=1thick,label=$n^{-1}$,label.side=left}
{pru,pru1}
\end{fmfgraph*}}\,
\nonumber\\ \nonumber\\ \nonumber\\ \nonumber\\ \nonumber
&+&\,\,\,\,\,\,\,\,\,
\parbox{30mm}{
\begin{fmfgraph*}(60,60)
\fmfleftn{l}{2} \fmfrightn{r}{4}
\fmfpoly{full}{ur,ul,uo}\fmfpen{thick} \fmfpoly{full}{dr,do,dl}
\fmfforce{(0.0w,0.2h)}{l1} \fmfforce{(0.0w,0.8h)}{l2}
\fmfforce{(1.2w,0.2h)}{r1}
\fmfforce{(1.2w,0.8h)}{r2}\fmfforce{(1.2w,1.1h)}{r3}\fmfforce{(1.2w,1.3h)}{r4}
\fmfforce{(0.9w,1.1h)}{o2}\fmfforce{(0.9w,0.9h)}{o1}
\fmfforce{(0.4w,0.8h)}{ul} \fmfforce{(0.6w,0.8h)}{ur}
\fmfforce{(0.5w,0.7h)}{uo} \fmfforce{(0.4w,0.2h)}{dl}
\fmfforce{(0.6w,0.2h)}{dr} \fmfforce{(0.5w,0.3h)}{do}
\fmf{fermion,width=1thick}{l2,ul}
\fmf{fermion,width=1thick,label=$n$,label.side=left}{ur,ddl}
\fmf{fermion,width=1thick,label=$p$,label.side=right}{ddr,r2}
\fmf{fermion,width=1thick}{l1,dl}
\fmf{fermion,width=1thick}{dr,r1}
\fmf{boson,width=1thick,label=$\pi^0$}{do,uo}
%\fmf{boson,width=1thick,label=$\pi^+$}{o1,uo}
\fmf{fermion,width=1thin}{r3,o2}
\fmf{fermion,width=1thin,label=$e$,label.side=left}{o2,r4}
\fmf{dashes,width=1thin}{o1,o2} \fmflabel{$\bar{\nu}$}{r3}
%\fmflabel{$e$}{r4}
%\fmflabel{$p$}{r2}
\fmflabel{$n$}{r1}\fmflabel{$n$}{l1}\fmflabel{$n$}{l2}
\fmfpoly{full}{ddr,o1,ddl}\fmfforce{(0.8w,0.8h)}{ddl}
\fmfforce{(1.0w,0.8h)}{ddr}
\end{fmfgraph*}}\,+\,\,...\,\,\,\,\,\,\,\,\,\\
\end{eqnarray}
It was shown in \cite{VS84}, (1986), (1987) and \cite{MSTV90}, see
also a more recent review \cite{V00}, that the main contribution
to the MMU process actually comes from the {\em{pion channel }} of
the reaction $nn\rightarrow npe\bar{\nu}$ (the first diagram
in (\ref{MMU-diag})), where $e\bar{\nu}$ are radiated from the
intermediate pion exchanging nucleons. Less contribution comes
from  the $NN^{-1}$ intermediate reaction states (second diagram),
and only much less contribution for $n \gsim n_0$ comes from the
nucleon of the leg of the reaction (third diagram, which naturally
generalizes the corresponding MU(FOPE) contribution (\ref{FOPEdiag})).

Moreover, due to the {\em{pion softening}} (medium modification of
the pion propagator) the matrix elements of the MMU process are
further enhanced with the increase of the density towards the pion
condensation critical point, see Fig. 1.
Roughly, the emissivity of MMU reaction then acquires  a factor
(mainly due to the pion decay channel of MMU)
\begin{eqnarray}\label{MMU}
\frac{\varepsilon_{\nu}[\rm MMU ]}{\varepsilon_{\nu}[\rm MU ]}
\sim 10^{3}~(n/n_0 )^{10/3}\frac{\Gamma^6 (n) }{[\omega^*  (n)/m_{\pi}]^8}
,
\end{eqnarray}
where the pre-factor $(n/n_0 )^{10/3}$ arises from the phase space
volume.

A different enhancement factor arises  for the MNB processes, where radiation
from intermediate reaction states (see first two diagrams in (\ref{MMU-diag})) is forbidden:
\begin{eqnarray}\label{MNB}
\frac{\varepsilon_{\nu}[\rm MNB ]}{\varepsilon_{\nu}[\rm NB ]}
\sim 10^{3}~(n/n_0 )^{4/3}\frac{\Gamma^6 (n) }{[\omega^* 
(n)/m_{\pi}]^3}.
\end{eqnarray}
The value $\omega^*$ entering eqs. (\ref{MMU}) and (\ref{MNB}) is determined
by curve 1a for $n<n_{c\pi}$ and by curve 2 for $n>n_{c\pi}$ if
a condensate is present. If a condensate is assumed to be absent one should use
the continuation 1b of the curve 1a.

\subsection{Medium effects in one-nucleon DU-like processes}
\subsubsection{MNPBF processes} The one-nucleon processes with
neutral currents given by the second diagram (\ref{nu-scat}) for
$N =(n, p)$ are forbidden at $T>T_{cN}$ by energy-momentum
conservation but they can occur at $T<T_{cN}$, where $T_{cN}$ is
the critical temperature for the nucleon-nucleon ($NN$) pairing. Then
the necessary energy and momentum can be taken from breaking and formation of the
Cooper pair.
However they need special techniques to be calculated, see
\cite{FRS76,VS87}. 
Their calculation is easily done in terms of closed 
diagrams with normal and anomalous Green functions, see \cite{VS87}. These diagrams
contain only one nucleon loop. It 
clearly demonstrates that these processes are indeed one-nucleon-like processes
rather than two-nucleon NB-like processes, as one sometimes interprets them. 
Due to the one-nucleon origin there appears a huge (for the gap $\Delta \gsim$~
MeV) pre-factor
$\sim 10^{29}$ in their emissivity, see eq. (\ref{PFB}) below.
Also the typical temperature pre-factor $\sim T^7$ 
following the rough estimate of \cite{FRS76}  is actually misleading. Instead
there is a $\Delta^7 (T/\Delta )^{1/2}$ pre-factor.
 
These processes (MnPBF) $n\rightarrow
n\nu\bar{\nu}$ and (MpPBF) $p\rightarrow p\nu\bar{\nu}$ play very
important roles in the cooling of superfluid NS, see
\cite{VS87,SV87,MSTV90,SVSWW97,BGV04}.  Due to the full vertices in (\ref{nu-scat})
extra $\Gamma_{w-s}^2$ factors appear in the MPBF emissivity.
Medium effects are not as
important for the $n\rightarrow n\nu\bar{\nu}$ process, changing
the emissivity by a factor of the order of one but they increase
the emissivity of the $p\rightarrow p\nu\bar{\nu}$ process by two
orders of magnitude, see eq. (\ref{p-vert}).

\subsubsection{Pion and kaon condensate processes} The $P$ wave pion
condensate can be of three types: $\pi_s^+$, $\pi^\pm$, and
$\pi^0$ with different values of the critical densities
$n_{c\pi}=(n_{c\pi^\pm}, n_{c\pi^+_{s}}, n_{c\pi^{0}}$), see
\cite{M78}. Thus above the threshold density for the pion
condensation of the given type, the neutrino emissivity of the MMU
process (\ref{MMU-diag}) is to be
supplemented by the corresponding PU processes\\
\\
\begin{eqnarray}\label{picond}
&&\parbox{30mm}{
\begin{fmfgraph*}(60,50)
\fmfleft{l} \fmfrightn{r}{3}
\fmfpoly{full}{ul,ur,uo}\fmfpen{thick} \fmfpoly{full}{dl,dr,do}
\fmfforce{(0.0w,0.6h)}{l} \fmfforce{(1.3w,0.6h)}{r1}
\fmfforce{(0.7w,0.9h)}{r2}\fmfforce{(0.7w,1.3h)}{r3}
\fmfforce{(0.95w,0.0h)}{o} \fmfforce{(0.4w,0.9h)}{o2}
\fmfforce{(0.4w,0.7h)}{uo} \fmfforce{(0.5w,0.6h)}{ur}
\fmfforce{(0.3w,0.6h)}{ul} \fmfforce{(0.95w,0.6h)}{dl}
\fmfforce{(1.15w,0.6h)}{dr} \fmfforce{(1.05w,0.5h)}{do}
\fmf{fermion,width=1thick}{l,ul}
\fmf{fermion,width=1thick,label=$p$}{ur,dl}
\fmf{fermion,width=1thick}{dr,r1}
\fmf{boson,width=1thick,label=$\pi^-_c $,label.side=left}{do,o}
%\fmf{boson,width=1thick,label=$\pi^+$}{o1,uo}
\fmf{fermion,width=1thin,label=$\bar{\nu}$,label.side=right}{r3,o2}
\fmf{fermion,width=1thin,label=$e$}{o2,r2}\fmf{dashes,width=1thin}{uo,o2}
%\fmflabel{$\bar{\nu}$}{r3}
\fmfv{decor.shape=cross,decor.size=4thick}{o}
\fmflabel{$n$}{r1}\fmflabel{$n$}{l}
\end{fmfgraph*}}\,\,\,\,\,\,\,\,\, , \,\,\,\,\,\,\,\,
\parbox{30mm}{
\begin{fmfgraph*}(60,50)
\fmfleft{l} \fmfrightn{r}{3}
\fmfpoly{full}{ul,ur,uo}\fmfpen{thick} \fmfpoly{full}{dl,dr,do}
\fmfforce{(0.0w,0.6h)}{l} \fmfforce{(1.3w,0.6h)}{r1}
\fmfforce{(0.7w,0.9h)}{r2}\fmfforce{(0.7w,1.3h)}{r3}
\fmfforce{(1.15w,0.0h)}{o} \fmfforce{(0.4w,0.9h)}{o2}
\fmfforce{(0.4w,0.7h)}{uo} \fmfforce{(0.5w,0.6h)}{ur}
\fmfforce{(0.3w,0.6h)}{ul} \fmfforce{(0.95w,0.6h)}{dl}
\fmfforce{(1.15w,0.6h)}{dr} \fmfforce{(1.05w,0.5h)}{do}
\fmf{fermion,width=1thick}{l,ul}
\fmf{fermion,width=1thick,label=$p$}{ur,dl}
\fmf{fermion,width=1thick}{dr,r1}
\fmf{boson,width=1thick,label=$\pi^+_c$,label.side=left}{do,o}
%\fmf{boson,width=1thick,label=$\pi^+$}{o1,uo}
\fmf{fermion,width=1thin,label=$\bar{\nu}$,label.side=right}{r3,o2}
\fmf{fermion,width=1thin,label=$e$}{o2,r2}\fmf{dashes,width=1thin}{uo,o2}
%\fmflabel{$\bar{\nu}$}{r3}
\fmfv{decor.shape=cross,decor.size=4thick}{o}
\fmflabel{$n$}{r1}\fmflabel{$n$}{l}
\end{fmfgraph*}}\,\,\,\,\,,
\,\,\,\,\,\,\,\nonumber\\\nonumber\\
\nonumber\\
&&\parbox{30mm}{
\begin{fmfgraph*}(60,50)
\fmfleft{l} \fmfrightn{r}{3}
\fmfpoly{full}{ul,ur,uo}\fmfpen{thick} \fmfpoly{full}{dl,dr,do}
\fmfforce{(0.0w,0.6h)}{l} \fmfforce{(1.3w,0.6h)}{r1}
\fmfforce{(0.7w,0.9h)}{r2}\fmfforce{(0.7w,1.3h)}{r3}
\fmfforce{(1.05w,0.0h)}{o} \fmfforce{(0.4w,0.9h)}{o2}
\fmfforce{(0.4w,0.7h)}{uo} \fmfforce{(0.5w,0.6h)}{ur}
\fmfforce{(0.3w,0.6h)}{ul} \fmfforce{(0.95w,0.6h)}{dl}
\fmfforce{(1.15w,0.6h)}{dr} \fmfforce{(1.05w,0.5h)}{do}
\fmf{fermion,width=1thick}{l,ul}
\fmf{fermion,width=1thick,label=$n$}{ur,dl}
\fmf{fermion,width=1thick}{dr,r1}
\fmf{boson,width=1thick,label=$\pi^0_c $,label.side=left}{do,o}
%\fmf{boson,width=1thick,label=$\pi^+$}{o1,uo}
\fmf{fermion,width=1thin,label=$\bar{\nu}$,label.side=right}{r3,o2}
\fmf{fermion,width=1thin,label=$e$}{o2,r2}\fmf{dashes,width=1thin}{uo,o2}
%\fmflabel{$\bar{\nu}$}{r3}
\fmfv{decor.shape=cross,decor.size=4thick}{o}
\fmflabel{$p$}{r1}\fmflabel{$p$}{l}
\end{fmfgraph*}}\,\,\,\, ,\,\,\,\,\,\,\,\, 
%%&&
\parbox{30mm}{
\begin{fmfgraph*}(60,50)
\fmfleft{l} \fmfrightn{r}{3}
\fmfpoly{full}{ul,ur,uo}\fmfpen{thick} \fmfpoly{full}{dl,dr,do}
\fmfforce{(0.0w,0.6h)}{l} \fmfforce{(1.3w,0.6h)}{r1}
\fmfforce{(0.7w,0.9h)}{r2}\fmfforce{(0.7w,1.3h)}{r3}
\fmfforce{(1.05w,0.0h)}{o} \fmfforce{(0.4w,0.9h)}{o2}
\fmfforce{(0.4w,0.7h)}{uo} \fmfforce{(0.5w,0.6h)}{ur}
\fmfforce{(0.3w,0.6h)}{ul} \fmfforce{(0.95w,0.6h)}{dl}
\fmfforce{(1.15w,0.6h)}{dr} \fmfforce{(1.05w,0.5h)}{do}
\fmf{fermion,width=1thick}{l,ul}
\fmf{fermion,width=1thick,label=$n$}{ur,dl}
\fmf{fermion,width=1thick}{dr,r1}
\fmf{boson,width=1thick,label=$\pi^0_c $,label.side=left}{do,o}
%\fmf{boson,width=1thick,label=$\pi^+$}{o1,uo}
\fmf{fermion,width=1thin,label=$\bar{\nu}$,label.side=right}{r3,o2}
\fmf{fermion,width=1thin,label=$\nu$}{o2,r2}\fmf{dashes,width=1thin}{uo,o2}
%\fmflabel{$\bar{\nu}$}{r3}
\fmfv{decor.shape=cross,decor.size=4thick}{o}
\fmflabel{$n$}{r1}\fmflabel{$n$}{l}
\end{fmfgraph*}}\,\,\,... \,\,\,\,\,\,\,\,\,\,
\end{eqnarray}
The wavy line with the cross is associated with the amplitude of the pion
condensate mean field. 
Contrary to the FOPE model, the MOPE model
of \cite{VS86} consistently takes into account the pion softening
effects for $n <n_{c\pi}$ and both the pion condensation and pion
softening effects in presence  of the condensate for $n
>n_{c\pi}$. As we have mentioned, in our numerical calculations
we assume  for simplicity that $n_{c\pi}=n_c^{\rm PU}\simeq 3~n_0$ is the same
for neutral and charged condensates. For vertices in (\ref{picond}) we use
the values presented above. Thus, emissivities of the PU processes are
suppressed
by the $\Gamma_s^2 \Gamma_{w-s}^2$ factors. In the case of a weak condensate
field
for non-superfluid matter all processes with charged
currents yield contributions to emissivity of the same order of magnitude,
whereas processes on neutral currents are significantly suppressed, see \cite{VS84},
(1986), \cite{L04}.

For $n >n_{c}^{\rm KU}$ the kaon condensate processes come into
play. The most popular is the idea of the $S$ wave $K^-$ condensation
(e.g. see \cite{BKPP88,T88}) which is allowed at $\mu_e
>m^*_{K^-}$ ($m^*_{K^-}$ is the effective kaon mass) due to possibility of the reaction $e\rightarrow
K^-\nu$. Analogous condition for the pion, $\mu_e >m^*_{\pi^-}$
($m^*_{\pi^-}$ is the effective
pion mass)
is not fulfilled due to a strong $S$ wave $\pi NN$ repulsion,
cf. \cite{M78,MSTV90}, (again the in-medium effect), otherwise $S$
wave $\pi^-$ condensation would occur at lower densities than
$K^-$ condensation. The neutrino emissivity of the $K^-$
condensate processes is given by the equation analogous to charged
pions however with a different $NN$ correlation factor and an additional
suppression factor due to a small contribution of the Cabibbo
angle.

The phase structure of dense NS matter might be very rich,
including $\pi^0$, $\pi^{\pm}$ condensates and $\bar{K}^0$,
$K^{-}$ condensates in both $S$ and $P$ waves, see \cite{KV03}; coupling
of condensates, see \cite{UNTMT94}; 
charged $\rho$-meson condensation, see \cite{V97,KV04}; fermion condensation yielding an
efficient DU-like process in the vicinity of the pion condensation
point (with the emissivity $\varepsilon_{\nu}\sim 10^{27}~T_9^5$,
$m^*_N \propto 1/T$), see \cite{VKZC00}; hyperonization, see
\cite{TT04}; quark matter with different phases, such as
2SC, CFL, CSL, plus their interaction with meson condensates, see
\cite{RW00,BGVquark1,BGVquark2} and refs therein; and different
mixed phases. In the present work, as in \cite{BGV04}, we suppress
all these possibilities of extra efficient cooling channels except
a PU process on the pion condensate. Other choices are
effectively simulated by our PU choice.

\subsubsection{Other resonance processes}
 There are many other
reaction channels allowed in the medium. Any Fermi liquid
permits propagation of zero sound excitations of different
symmetry related to the pion and the quanta of a more  local
interaction determined via Landau-Migdal parameters $f_{\alpha ,\beta}$ and $g_{\alpha
,\beta}$. These excitations present at $T\neq 0$ may also
participate in the neutrino reactions. The most essential
contribution comes from the neutral current processes, see
\cite{VS86}, given by the first two diagrams of the
series\\
\begin{eqnarray}\label{res-pr}
\parbox{30mm}{
\begin{fmfgraph*}(60,60)
\fmfpen{thick}
\fmfforce{(0.7w,1.0h)}{r3}\fmfforce{(0.3w,1.0h)}{r4}
\fmfforce{(0.5w,0.8h)}{o2}\fmfforce{(0.5w,0.65h)}{o1}
\fmfforce{(0.0w,0.5h)}{uo}\fmfforce{(0.3w,0.5h)}{o4}
\fmfforce{(1.0w,0.5h)}{do}\fmfforce{(0.7w,0.5h)}{o3}
\fmf{dots,width=1thick}{do,o3} \fmf{dots,width=1thick}{o4,uo}
\fmf{fermion,width=1thin}{o2,r3}
\fmf{fermion,width=1thin}{o2,r4}\fmf{dashes,width=1thin}{o1,o2}
\fmflabel{$\nu$}{r3}\fmflabel{$\bar{\nu}$}{r4}
\fmfpoly{full}{o1,o5,o6}
\fmfforce{(0.4w,0.6h)}{o5}\fmfforce{(0.6w,0.6h)}{o6}
\fmfpoly{full}{or,pru,prd} \fmfforce{(0.7w,0.5h)}{or}
\fmfforce{(0.6w,0.6h)}{prd} \fmfforce{(0.6w,0.4h)}{pru}
\fmfpoly{full}{o4,pru1,prd1} \fmfforce{(0.3w,0.5h)}{o4}
\fmfforce{(0.4w,0.6h)}{prd1} \fmfforce{(0.4w,0.4h)}{pru1}
\fmf{fermion,left=.5,tension=.5,width=1thick} {prd1,prd}
\fmf{fermion,left=.5,tension=.5,width=1thick,label=$n^{-1}$,label.side=left}
{pru,pru1}
\end{fmfgraph*}}\,+\,\,\,\,\,\,\,
\parbox{30mm}{
\begin{fmfgraph*}(60,50)
\fmfleft{l} \fmfrightn{r}{3}
\fmfpoly{full}{ul,ur,uo}\fmfpen{thick} \fmfpoly{full}{dl,dr,do}
\fmfforce{(0.0w,0.8h)}{l} \fmfforce{(1.3w,0.8h)}{r1}
\fmfforce{(0.7w,1.1h)}{r2}\fmfforce{(0.7w,1.3h)}{r3}
\fmfforce{(1.05w,0.2h)}{o} \fmfforce{(0.4w,1.1h)}{o2}
\fmfforce{(0.4w,0.9h)}{uo} \fmfforce{(0.5w,0.8h)}{ur}
\fmfforce{(0.3w,0.8h)}{ul} \fmfforce{(0.95w,0.8h)}{dl}
\fmfforce{(1.15w,0.8h)}{dr} \fmfforce{(1.05w,0.7h)}{do}
\fmf{fermion,width=1thick}{l,ul}
\fmf{fermion,width=1thick,label=$n$}{ur,dl}
\fmf{fermion,width=1thick}{dr,r1} \fmf{dots,width=1thick}{do,o}
%\fmf{boson,width=1thick,label=$\pi^+$}{o1,uo}
\fmf{fermion,width=1thin}{r3,o2}
\fmf{fermion,width=1thin,label=$\nu$}{o2,r2}\fmf{dashes,width=1thin}{uo,o2}
\fmflabel{$\bar{\nu}$}{r3}
%\fmfv{decor.shape=cross,decor.size=4thick}{o}
\fmflabel{$n$}{r1}\fmflabel{$n$}{l}
\end{fmfgraph*}}\,\,\,\,\,\,+\,\,\, ...
\end{eqnarray}
Here the dotted line is the zero sound quantum of the appropriate
symmetry. These are the resonance processes (second one of the DU-type)
analogous to those processes on the condensates with the
only difference that the rates of reactions with zero sounds are
proportional to the thermal occupations of the corresponding
spectrum branches whereas the rates of the condensate processes
are proportional to the modulus squared of the condensate mean field.
The contribution of the resonance reactions is  rather small
due to the small phase space volume ($q\sim T$) associated with zero
sounds. Note the analogy of the processes
(\ref{res-pr}) with the corresponding phonon processes in the crust.

\subsubsection{DU processes}
 The proper DU processes in matter, as
$n\rightarrow pe^-\bar{\nu}_e$ and $ pe^-\rightarrow n\nu_e$,
\begin{eqnarray} \label{du}
\,\,\,\,\,\,\,\,\,\,\,\,\,\,\,\,\,\,\,\,\,\,\,\,
\parbox{10mm}{\begin{fmfgraph*}(50,35)
\fmfpen{thick} \fmfleft{l1} \fmfright{r1,r2,r3}
\fmf{fermion}{l1,T2} \fmf{fermion}{T3,r1}
\fmf{fermion,width=1thin}{o3,r3}\fmf{fermion,width=1thin}{r2,o3}
\fmf{dashes,width=1thin}{T1,o3}
%%%\fmfpoly{shaded,width=1thin}{p1,p2,p3,p4}
%%%\fmfblob{.05w}{o1}
\fmfpoly{full}{T2,T3,T1} \fmflabel{$e$}{r3}
\fmflabel{$\bar{\nu}$}{r2}\fmflabel{$n$}{l1}\fmflabel{$p$}{r1}\end{fmfgraph*}}
\,\,\,\,\,\,\,\,\,\,\,\,\,\,\,\,\,\,\,\,\,\,\,\,\,\,+
\,\,\,\,\,\,\,\,\,\,\,\,
\parbox{10mm}{\begin{fmfgraph*}(50,35)
\fmfpen{thick} \fmfleft{l1,l2} \fmfright{r1,r2}
\fmf{fermion}{l1,T2} \fmf{fermion}{T3,r1}
\fmf{fermion,width=1thin}{l2,o3}\fmf{fermion,width=1thin}{o3,r2}
\fmf{dashes,width=1thin}{T1,o3}
%%%\fmfpoly{shaded,width=1thin}{p1,p2,p3,p4}
%%%\fmfblob{.05w}{o1}
\fmfpoly{full}{T2,T3,T1} \fmflabel{$e$}{l2}
\fmflabel{$\nu$}{r2}\fmflabel{$p$}{l1}\fmflabel{$n$}{r1}\end{fmfgraph*}}
\,\,\,\,\,\,\,\,\,\,\,\,\,\,\,\,\,\,\,\,\,\,\,\,\,
\end{eqnarray}\\
should also be treated with the full vertices.  They are forbidden
up to the density $n_c^{\rm DU}$ when the triangle inequality
$p_{Fn}<p_{Fp}+p_{Fe}$ begins to be fulfilled. For traditional EoS like
that given by the variational theory by \cite{APR98},  DU processes
are permitted only for $n >5\,n_0$. Due to the full vertices in (\ref{du})
extra $\Gamma_{w-s}^2$ factors appear in the DU emissivity.

\section{Gaps}\label{gaps}

%%\subsection{Nucleon superfluidity}

\begin{figure}[htb]
\psfig{figure=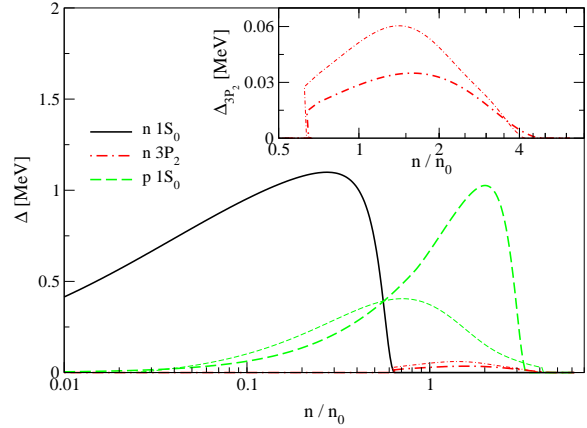,width=0.5\textwidth,angle=-90} \caption{
Neutron and proton pairing gaps according to model I
%%\cite{YGKLP03}
(thick solid,
dashed and dotted lines) and according to model II
%%\cite{TT04}
%%%{SVSWW97}
(thin lines), see text. The $1S_0$ neutron gap is the same in both
models, taken from  \cite{AWP}. \label{fig-gaps}}
\end{figure}

In spite of the many calculations that have been performed, the
values of nucleon gaps in dense NS matter are poorly known. This
is the consequence of the exponential dependence of the gaps on
the density dependent potential of the in-medium $NN$ interaction.
This potential is not sufficiently well known. Gaps that we have
adopted in the framework of the ''nuclear medium cooling''
scenario, see \cite{BGV04}, are presented in Fig. \ref{fig-gaps}.
Thick dashed lines show proton gaps used in the work of
\cite{YGKLP03} performed in the framework of the ``standard plus
exotics'' scenario. In their model, proton gaps are artificially
enhanced (that is, not supported by  any microscopic calculations)
to get a better fit of the data. We use their ``1p'' model.
Neutron $3P_2$ gaps presented in Fig. \ref{fig-gaps} (thick dash
-dotted lines) are the same as those of the ``3nt'' model of
\cite{YGKLP03}. We will call this model I. Thin lines
show $1S_0$ proton and $3P_2$ neutron gaps from \cite{TT04}, for
the model AV18 by \cite{WSS95} (we call it model II). We take
the same $1S_0$ neutron gap  in both models I and II (thick solid
line), as  calculated  by \cite{AWP} and used
by \cite{SVSWW97} within the cooling code.
 \cite{BGV04} used  both models I and II
within the ``nuclear medium cooling'' scenario. Since the  $1S_0$ neutron pairing gap exists only within
the crust, dying for baryon densities $n\geq 0.6~n_0$, its effect
on the cooling is rather minor. The effect on the
cooling arising from the proton  $1S_0$ pairing and  from the
neutron $3P_2$ pairing, with gaps reaching up to rather high
densities, is pronounced. The NS cooling essentially depends on
the values of the gaps and on their density dependence. Findings
of \cite{SCLBL96} and \cite{LS00}, who incorporated in-medium effects,
motivated us to check the possibility of a suppression of  $1S_0$
neutron and proton gaps. For that we  introduced pre-factors
for $1S_0$ neutron and proton gaps which we varied in the range
$0.2\div 1$, see Figs. 18 and 19 of \cite{BGV04}.
%%%%%%%%%%%%%%%%%%%% Figure 5 %%%%%%%%%%%%%%%%%%
%%%%%%%%%%%%%%%%%%%%%%%%%%%%%%%%%%%%%%%%%%%%%%%%

\subsection{Possibilities either of strong suppression or strong enhancement
  of $3P_2$
gap}
Recently \cite{SF03} have argued for a   strong suppression of the
$3P_2$ neutron gaps, down to values $\lsim 10~$keV, as a
consequence of the  medium-induced spin-orbit interaction. They
included important medium effects, such as the modification of the
effective interaction of particles at the Fermi surface owing to
polarization contributions, with particular attention to
spin-dependent forces. In addition to the standard spin-spin,
tensor and spin-orbit forces, spin non-conserving effective
interactions were induced by screening in the particle-hole
channels. Furthermore a novel long-wavelength tensor force was
generated. The polarization contributions were computed to second
order in the low-momentum interaction $V_{\rm{low}\,k}$. These
findings motivated \cite{BGV04} to suppress values of $3P_2$ gaps
shown in Fig. \ref{fig-gaps} by an extra factor $f(3P_2 ,n)=0.1$.
Further possible suppression of the $3P_2$ gap is almost not
reflected in the behavior of the cooling curves.

Contrary to expectations of \cite{SF03}, the  more recent
work of \cite{KCTZ04} argued that the
%%$^3$P$_2$
$3P_2$ neutron pairing gap should be
dramatically enhanced,
as a consequence of  the strong softening of the pion propagator. 
According
to their estimation,
the
%%$^3$P$_2$
$3P_2$ neutron pairing gap is  as large as $1\div 10$~MeV in a
broad region of densities, see Fig. 1 of their work.

In order to apply these results to a broad density
interval both models may need further improvement. The model  of
\cite{SF03} was developed to describe not too high densities. It
does not incorporate higher order nucleon-nucleon hole loops and
the $\Delta$ isobar contributions and thus it may only partially
include the pion softening effect at densities $\gsim n_0$.
The model of \cite{KCTZ04} uses a  simplified analytic
expression for the effective pion gap $(\omega^*)^2 (k_m)$, valid near the pion condensation
critical point, if the latter occurred by a second order phase
transition. The latter assumption means that $(\omega^*)^2
(k_m)$ is assumed to be zero in the critical point of the phase
transition. Outside the vicinity of the critical point the
parameterization  of the effective pion gap that was used can be
considered only as a rough interpolation. Actually the phase
transition is of first order and evaluations of quantum
fluctuations done by
 \cite{D82}
show that the value of the jump of the effective pion gap  in the
critical point is not as small. Moreover  repulsive  correlation
contributions to the $NN$ amplitude have been disregarded. In the
pairing channel under consideration, already outside  the narrow
vicinity of the pion condensation critical point, the repulsion
originating from the $NN$ correlation effects may exceed the
attraction originating from  the pion softening. Notice that, if
the pairing gap enhancement occurred only in a rather narrow
vicinity of the pion condensation critical point, it would not
affect the results of \cite{BGV04}. In the latter work two
possibilities were considered (see Fig. 1): i) a saturation of the
pion softening with increase of the baryon density resulting in
the absence of  pion condensation and ii) a stronger pion
softening stimulating the occurrence of the pion condensation for
$n>n_c \simeq 3~n_0$. In both cases the effective pion gap was
assumed never to approach zero and undergoing a not too small jump
at the critical point from a finite positive value
($(\omega^*)^2 \simeq 0.3~m_{\pi}^2$) to a finite negative value ($(\omega^*)^2
\simeq -0.1~m_{\pi}^2$). The reason for such a strong jump is the
strong coupling. If it were so, the strong softening assumed by
\cite{KCTZ04} would not be realized. However, due to uncertainties
in the knowledge of forces acting in strong interacting nuclear
matter and a poor description of the vicinity of the phase
transition point we cannot exclude that the alternative possibility
of a tiny jump of the pion gap exists. Therefore we will check how
these  alternative hypotheses may work within our "nuclear medium
cooling" scenario. Avoiding further discussion of the
theoretical background of the models,  we investigate
the possibility of a significantly enhanced
$3P_2$ neutron
pairing gap and of a partially suppressed proton $1S_0$ gap, as 
suggested by  \cite{KCTZ04}. To proceed  in the framework of our
 ``nuclear medium cooling'' scenario
we introduce the enhancement factor of the original $3P_2$ neutron
pairing gap $f (3P_2 ,n)$, and a suppression factor of the proton $1S_0$
gap $f(1S_0 ,p)$.  We do not change the neutron $1S_0$ gap since  its effect on the cooling is minor.

\subsection{Suppression of the emissivity by superfluid gaps}

 Generally speaking, the
suppression factors of superfluid processes are given by
complicated integrals. As was demonstrated  by \cite{Sedr04} with
the example of the DU process, these integrals are not
reduced to the  $R$-factors, see \cite{YGKLP03}. However,
for temperatures significantly below the critical temperature the
problem is simplified. With an exponential accuracy the
suppression of the specific heat is governed by  the factor
$\xi_{nn}$ for neutrons and $\xi_{pp}$ for protons:
\begin{eqnarray}
\xi_{ii}\simeq  \mbox{exp}[{-\Delta_{ii}(T)/T}], \quad \mbox{for}\quad T<T_{ci};\quad i=n,p,
\end{eqnarray}
and $\xi_{ii}=1$ for $T>T_{ci}$, $T_{ci}$ is the corresponding critical
temperature.
We do not need  higher accuracy to
demonstrate our result. Therefore we will
use these simplified factors.

For the emissivity of the DU process the suppression factor is given by
$\mbox{min}\{\xi_{nn},\xi_{pp}\}$, see \cite{LP}. The same suppression factor
appears for other one nucleon processes, as PU and KU and the resonance
processes on zero sounds, see the second diagram of (\ref{res-pr}).
Suppression factors for two nucleon processes follow from this fact and from
the diagrammatic representation of different processes within
the closed diagram technique by \cite{VS87} and Knoll \& Voskresensky  (1995, 1996). These are:
 $\xi_{nn}\cdot \mbox{min}\{\xi_{nn},\xi_{pp}\}$
for the neutron branch of the  MU  process (and for the 
same branch of the 
medium
modified Urca process, MMU); $\xi_{pp}\cdot
\mbox{min}\{\xi_{nn},\xi_{pp}\}$ for the  corresponding proton
branch of the process; $\xi_{nn}^2$ for the  neutron branch of the
(medium modified) nucleon bremsstrahlung (MnB) and $\xi_{pp}^2$
for the corresponding  proton branch of the bremsstrahlung (MpB).
Thus, for $\Delta_{nn}\gg \Delta_{pp}$ both neutron and proton
branches of the MMU process  are frozen for $T\ll T_{cn}$  due to
the factors $\xi_{nn}^2$ and $\xi_{pp}\xi_{nn}$, respectively.
Zero sound and other phonon processes shown by the first diagram
(\ref{res-pr})
are not suppressed by the $\xi_{ii}$ factors. However their contribution to
the emissivity is
very small due to the smallness of the available phase space volume.

\section{Cooling model of \cite{BGV04}}
\subsection{EoS and structure of NS interior, crust,
surface}\label{reg} \subsubsection{NS interior} We will exploit
the EoS of \cite{APR98} (specifically the Argonne $V18+\delta
v+UIX^*$ model), which is based on the most recent models for the
$NN$ interaction with the inclusion of a parameterized
three-body force and relativistic boost corrections. Actually we
adopt a simple analytic parameterization of this model given by
\cite{HJ99} (HHJ). It uses the compressional part with the
compressibility $K\simeq 240$~MeV, and a symmetry energy fitted to
the data around the nuclear saturation density, and smoothly
incorporates causality at high densities. The density dependence
of the symmetry energy is very important since it determines the
value of the threshold density for the DU process. The HHJ EoS
fits the symmetry energy  to the original Argonne $V18+\delta v
+UIX^*$ model yielding $n_c^{\rm DU}\simeq~5.19~n_0$ ($M_c^{\rm
DU}\simeq 1.839~M_{\odot}$).  The original Argonne EoS allows
for  neutral pion condensation (for $n>2n_0$)  that only
slightly affects the energy density. One may disregard this small
change.
This EoS does not allow for  charged pion condensation. 
The HHJ parameterization of the EoS does not include $\pi$ condensation effects.
We further assume that  pion condensation (neutral and charged) occurs for $n>3n_0$,
see discussion in \cite{BGV04}. We assume a minor effect of the pion condensation
on the EoS and disregard it.  We also disregard changes of the isotopic composition
due to the charged pion
condensation. The latter effect would be small only if the charged pion condensate field
were rather weak.
Thus we
assume that  the HHJ parameterization of the  EoS 
includes both mentioned effects or that they are negligible.

\subsubsection{NS crust} The density $n\sim 0.5\div 0.7 ~n_0$ is
the boundary of the NS interior and the inner crust. In the latter there
occurs the
so-called ``pasta phase'' discussed by \cite{RPW83}, see also
 \cite{MTVTCM04}. Then there is the outer crust and
the envelope. Note that our code generates a  temperature
profile  that inhomogeneous during the first $10^2 \div 10^3~$ y. The
influence of the crust on the cooling and heat transport is 
minor  due to its rather low mass content.  Therefore the
temperature also changes slightly in the crust up to the envelope.

\subsubsection{Envelope}

\begin{figure}[htb]
\psfig{figure=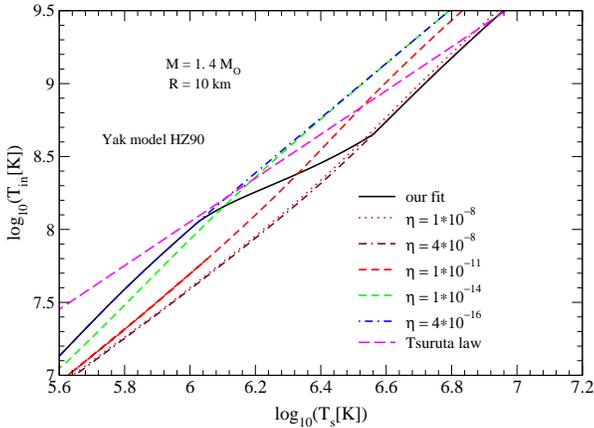,width=0.5\textwidth,angle=-90}
\caption{The relation between the inner crust temperature and the
surface temperature for different models. Dash-dotted curves
indicate boundaries of the uncertainty band. Notations of lines
are determined in the legend. For more details see  \cite{BGV04}
and \cite{YLPGC03}. \label{T-in} }
\end{figure}

The resulting cooling curves depend on the $T_{\rm in}-T_{\rm s}$
relation between internal and surface temperatures in the
envelope. Fig. \ref{T-in} shows uncertainties existing in this
relation. 
%%%%%%%%%%%%%%%%%%%% Figure 4 %%%%%%%%%%%%%%%%%%
%%%%%%%%%%%%%%%%%%%%%%%%%%%%%%%%%%%%%%%%%%%%%%%%
A calculation is presented for  the canonical NS: $M=1.4
M_{\odot}$, $R=10$~km with the crust model HZ90 of \cite{YLPGC03}.
Below we will show that a minimal discrepancy of the increased gap scenario
with the data is
obtained with what we called ``our fit'' model in \cite{BGV04}. Using other
choices like the ``Tsuruta law" ($T_{\rm s}^{\rm Tsur}=(10 T_{\rm
in})^{2/3}$, where $T_{\rm s}$ and $T_{\rm in}$ are measured in K)
only increases the discrepancy. To compare results with ``our
fit'' model we  use the upper boundary curve, $\eta =4\cdot
10^{-16}$\, and the lower boundary curve $\eta =4\cdot
10^{-8}$\,. In Fig.\ref{T-in} we also draw lines $\eta
  =1\cdot 10^{-14}$ and $\eta =1\cdot 10^{-11}$ as they are
  indicated in the corresponding Fig. 2 of \cite{YLPGC03}.
The selection of $\eta =4\cdot 10^{-8}$
and $\eta =4\cdot 10^{-16}$ as the boundaries of the
uncertainty-band seems to be a too strong restriction, see
\cite{YLPGC03}. The limit of the most massive helium layer is
achieved for
 $\eta \sim
10^{-10}$.
On the other hand the
 helium layer begins to affect the thermal structure
only for $\eta >
10^{-13}$. Thus one could exploit $10^{-13}<\eta < 10^{-10}$ as a $T_{\rm
  in}-T_{\rm s}$ band.
  We will use a broader band, as is shown in
Fig.\ref{T-in}. By this we simulate the effect of maximum
uncertainties in the knowledge of the $T_{\rm  in}-T_{\rm s}$
relation.

\subsection{Main cooling regulators} We compute the NS thermal
evolution adopting our fully general relativistic evolutionary
code. This code was originally constructed for the description of
hybrid stars by \cite{BGVquark1}. The main cooling regulators are
the thermal conductivity, the heat capacity and the emissivity. In
order to better compare our results with results of other groups
we try to be as close as possible to their inputs for the
quantities that we did not calculate ourselves. Then we add
the changes, improving the EoS and including the medium effects.

\subsubsection{Thermal conductivity}
We take the electron-electron
contribution to the thermal conductivity and the electron-proton
contribution for normal protons from \cite{GY95}. The total
contribution related to electrons is then given by
\begin{eqnarray}\label{kapel}
1/\kappa_e = 1/\kappa_{ee}+1/\kappa_{ep}.
\end{eqnarray}
For $T>T_{cp}$ (normal "n" matter), we have $\kappa_{ep}^{\rm n}
=\kappa_{ep}$. For $T<T_{cp}$ (superfluid "s" matter) we use the
expression
\begin{eqnarray}
\kappa_{ep}^{\rm s}  = \kappa_{ep}/\xi_{pp} >  \kappa_{ep}^{\rm n} ,
\end{eqnarray}
that gives a crossover from the non-superfluid case to the
superfluid case. The vanishing of $\kappa_{ep}^{\rm s}$ for $T\ll
T_{cp}$ is a consequence of the scattering of superfluid protons
on the electron impurities, see \cite{BGVquark1}. Following
(\ref{kapel}) we get $\kappa_e^{\rm n}  < \kappa_e^{\rm s}$.

For the neutron contribution,
\begin{eqnarray}
\kappa_n = 1/\kappa_{nn}+1/\kappa_{np}~,
\end{eqnarray}
we use the result of \cite{BHY01} that includes corrections due to
the superfluidity. Although some medium effects are incorporated
in this work, the nucleon-hole corrections of correlation terms
and the modification of the tensor force are not included. This
should modify the result. However, since we did not calculate
$\kappa_n$ ourselves, we can only roughly estimate the
modification: not too close to
the critical point of the pion condensation the squared matrix
element of the $NN$ interaction  $|M|^2_{\rm med} \sim p_{F,n}^2
\Gamma^2_s /(\omega^*)^2$, see eq. (\ref{amplMOPE}) and values shown in Fig.1, 
is of the order of the
corresponding quantity
 $|M|^2_{\rm vac} \sim p_{F,n}^2 /[m_{\pi}^2 + p_{F,n}^2 ]$
estimated with the free one pion exchange, whereas $|M|^2_{\rm
med}$ may significantly increase for $n\sim n_c^{\rm PU}$.
However \cite{BGV04} checked that both increasing and decreasing of the
thermal conductivity does not change the picture as a whole.

The proton term is calculated similarly to the neutron one,
\begin{eqnarray}\label{p}
\kappa_p = 1/\kappa_{pp}+1/\kappa_{np}~.
\end{eqnarray}

 The total thermal conductivity is the  sum of the
partial contributions
\begin{eqnarray}
\kappa_{tot} = \kappa_{e}+\kappa_{n}+\kappa_p +...
\end{eqnarray}
For the values of the gaps used in \cite{BGV04} the other
contributions to this sum are smaller than those presented
explicitly ($\kappa_e$, $\kappa_n$ and $\kappa_p$). Finally
\cite{BGV04} concluded that in their scenario transport is
relevant only up to the first $ 10^3$~y.

\subsubsection{Heat capacity} 
The heat capacity contains nucleon,
electron, photon, phonon and other contributions. The main
in-medium modification of the nucleon heat capacity is due to the
density dependence of the effective nucleon mass. We use the same
expressions as \cite{SVSWW97}. The main regulators are the nucleon
and the electron contributions. For the nucleons ($i=n,p$), the
specific heat is (\cite{M79})
\begin{equation}\label{spec}
c_i \sim 10^{20}({m_i^*}/{m_N})~(n_i/n_0)^{1/3} \xi_{ii}~T_9~
{\rm erg~cm^{-3}K^{-1}}~,
\end{equation}
for the electrons it is
\begin{equation}\label{e}
c_e \sim 6\times 10^{19} \,(n_e/n_0)^{2/3}~T_9~ {\rm
erg~cm^{-3}~K^{-1}}~ .
\end{equation}
%%where $Y_e=n_e/n$.
Near the phase transition point the heat capacity acquires a
pion fluctuation contribution. For the first order pion condensation
phase transition this additional contribution contains no
singularity, like for the second order
phase transition, see \cite{VM82} and \cite{MSTV90}. Finally, the nucleon
contribution to the heat capacity may increase up to several times
in the vicinity of the pion condensation point. The effect of this
correction on global cooling properties is unimportant
and simplifying we neglect it.

The symmetry of the $3P_2$ superfluid phase allows for the
contribution of Goldstone bosons (phonons):
\begin{eqnarray}\label{phon}
C_{G}\simeq 6\cdot 10^{14}T_9^3 \,\,\, \frac{{\rm erg}}{{\rm
cm}^{3}~{\rm K}},
\end{eqnarray}
for $T<T_{cn}(3P_2)$, $n>n_{cn}(3P_2)$. We  include this
term in our study although its effect on the cooling is
minor. A similar contribution comes from other resonance processes
permitted on zero sounds, see the first diagram in (\ref{res-pr}). In order not to
introduce an extra parameter dependence we will simulate the  effect of all phonon
and zero sound terms with the term (\ref{phon}).

\subsubsection{Emissivity} We adopt the same set of partial
emissivities as in the work of \cite{SVSWW97}. The phonon
contribution to the emissivity of the $3P_2$ superfluid phase, 
as well as the zero sound contribution, is
negligible. The main emissivity regulators are the MMU, see the
rough estimation (\ref{MMU}), MnPBF and MpPBF, and MNB processes.
For $n>n_c^{\rm PU}\simeq 3~n_0$ the PU process becomes efficient and for
$n>n_c^{\rm DU}\simeq 5.19~n_0$
the DU process is the dominant.

Only the qualitative behavior of the interaction shown in  Fig.
\ref{fig1} is motivated by the microscopic analysis whereas the actual
numerical values of the correlation parameter and the pion gap are
rather uncertain. Thus we vary the values $\Gamma (n)$ and
${\omega^*}^2 (n)$ in accordance with the discussion of Fig.
\ref{fig1}. By that we check the relevance of alternative
possibilities: a) no pion condensation and a saturation of the
pion softening with increasing density, curves 1a$+$1b, and b) the presence of pion
condensation, curves 1a$+$2$+$3. We also add the contribution of the DU for
$n>n_c^{\rm DU}$.

All emissivities are corrected by correlation effects. The PU
process contains an extra $\Gamma_s^2$ factor compared to the DU
process (the emissivity of the latter is $\propto \Gamma_{w-s}^2$).
 Another suppression of PU emissivity comes from the fact
that it is proportional to the squared pion condensate mean field
$|\varphi|^2$. Near the critical point $|\varphi|^2 \sim 0.1$,
increasing with  density up to $|\varphi|^2 \sim f_{\pi}^2 /2$,
where $f_{\pi}\simeq 93~$MeV is the pion decay constant. Finally,
the PU emissivity is suppressed by  about 1-2 orders of magnitude
compared to the DU one. We adopt the same gap dependence
for the PU process as  for the corresponding DU process. In superfluid
matter all emissivity terms are suppressed by the corresponding $\xi_{ii}$ factors.

\section{Main cooling regulators in the scenario of
\cite{KCTZ04}}\label{Khodel}

When neutron processes are frozen the most efficient
process is the MpPBF process, $p\rightarrow p\nu\bar{\nu}$, for
$T<T_{cp}$. Taking into account the medium effects in the weak
coupling vertex, see eq. (\ref{p-vert}),  we use the same expression for the emissivity of
this process as has been used by \cite{V00} and \cite{BGV04}:
\begin{eqnarray}\label{PFB}
&&\varepsilon_{\nu} [\mbox{MpPBF}]\sim 10^{29}~ \frac{m_N^*}{m_N}
\left[\frac{p_{Fp}}{p_{Fn}(n_0)}\right]
~\left[\frac{\Delta_{pp}}{\mbox{MeV}}\right]^{7}~
\nonumber\\
&&\times \left[\frac{T}{\Delta_{pp} }\right]^{1/2}~\xi_{pp}^2 ~~\frac{{\rm
    erg}}{{\rm cm}^{3}~{\rm sec}}  ~, \quad T<T_{cp}.
\end{eqnarray}
This process contributes only below the critical
temperature for proton pairing. Inclusion of medium effects
greatly enhances the vertex of this process compared to the vacuum
vertex, see the above discussion of this fact in subsection \ref{ren}. 
A factor $\Gamma_{w-s}^2\sim 10^2$ arises, since the process
may occur through $nn^{-1}$ and $ee^{-1}$ correlation states, with subsequent production
of $\nu\bar{\nu}$ from the $nn^{-1}\nu\bar{\nu}$ and
$ee^{-1}\nu\bar{\nu}$ channels rather than from a strongly
suppressed channel $pp^{-1}\nu\bar{\nu}$, see
\cite{VS87,SV87,MSTV90,VKK98,L00} and \cite{V00}. Relativistic corrections
incorporated in 
the description of the $pp^{-1}\nu\bar{\nu}$
vertex  in \cite{YGKLP03}
also produce an enhancement being however significantly less than
that arises from medium effects in $nn^{-1}\nu\bar{\nu}$ and
$ee^{-1}\nu\bar{\nu}$ channels. Thus we see no
reason not to include these medium effects and we note
only a moderate dependence  of the result on the uncertainties in
the strong interaction.

We also present here an explicit expression for the emissivity of
the proton branch of the nucleon bremsstrahlung including medium
effects, MpB, $pp\rightarrow pp\nu\bar{\nu}$. In the case of
suppressed neutron $3P_2$ gaps this process contributes much less
than the nucleon bremsstrahlung processes involving neutrons. However, when neutron reactions are
frozen, the $pp\rightarrow pp\nu\bar{\nu}$ process becomes the
dominating process for $T_{cn}>T>T_{cp}$. The emissivity of the
$pp\rightarrow pp\nu\bar{\nu}$ reaction takes the form, see
\cite{VS86} for more details,
\begin{eqnarray}\label{MpB}
&&\epsilon({\rm MpB})\sim  10^{23}\xi_{pp}^2  I_{pp}\frac{Y_p^{5/3}
 \Gamma_s^4 T_9^8}{(\omega^*)^4 (k_m)}\nonumber\\
&&\times\left(\frac{m_{p}^*}{m_N}\right)^4
\left(\frac{n}{n_0}\right)^{5/3}~~\frac{{\rm
    erg}}{{\rm cm}^{3}~{\rm sec}},
\end{eqnarray}
%%in $\mbox{erg}/(\mbox{cm}^3\cdot \mbox{sec})$,
$T_9 =T/10^9$ K, $m_p^* $ is the effective proton
 mass. Here we take 
 $\Gamma_s \simeq \Gamma (n)\simeq 1/[1+C(n/n_0)^{1/3}]$, $C\simeq 1.4\div
 1.6$. This factor takes into
 account $NN$ correlations
in strong interaction vertices,
 $Y_p =n_p/n$ is the proton to nucleon ratio.
As above, for simplicity  we assumed that the value $k=k_m$ (at which the
square of the effective pion gap $(\omega^*)^2 (k)$ is minimimum) is
rather close to the value of the neutron Fermi momentum $ p_{
Fn}$, as it follows from the microscopic analysis
of \cite{MSTV90}.
To simplify the consideration we take the same value of the effective
pion gap for the given process  as that for the MMU
 process (although in general  it is not so, and thus the result
 (\ref{MpB}) is model dependent),
cf. \cite{BGV04},
\begin{eqnarray}\label{intI}
I_{pp} \sim \frac{\pi}{64}
\left(\frac{p_{Fn}}{p_{Fp}}\right)^5 \frac{\omega^* (k_m)}{p_{Fn}}.
\end{eqnarray}
We have checked that for $T<T_{cp}$ for the pairing gaps under
consideration the MpB reaction contributes significantly less than
the MpPBF process. It could be not the case only  in a narrow
vicinity of the pion condensation critical point, if pion
condensation occurred with only a tiny jump of the effective pion
gap in the  critical point. However, even in this case there are
many effects which could mask this abnormal enhancement.

For $n>n_c^{\rm PU}$ the process $p\pi^0_c \rightarrow p\nu\bar{\nu}$ on
the neutral pion condensate is still permitted. However its contribution is 
strongly suppressed, see \cite{L04}.

In the case of frozen neutron degrees of freedom the specific heat is governed by
protons and electrons,
see eq. (\ref{spec}) for $i=p$ and (\ref{e}).
Here, we again suppress the contribution to the specific heat by the 
narrow vicinity of the pion condensation critical point  due to
the fact that in our scenario (see Fig. 1) the modulus of the
squared effective pion gap $(\omega^*)^2 $ is always larger than
$\sim (0.1\div 0.3)~m_{\pi}^2$. With such an effective pion gap
the pion contribution to the specific heat  is not as strong and
can be disregarded to simplify the consideration. For the
second order phase transition (and for a first order phase
transition but with a tiny jump of $|(\omega^*)^2 |$ in
the critical point), pion fluctuations would contribute  more strongly
to the specific heat yielding a term $c_{\pi} \propto
T^{\rm PU}_c /\omega^*$ for $|T- T^{\rm PU}_c |/ T^{\rm PU}_c\ll T^{\rm PU}_c$, see \cite{VM82,MSTV90}.

When neutron processes are frozen the values $\kappa_n$
and $\kappa_p$ are  suppressed and the thermal conductivity is
reduced to the electron and proton contributions, see
eqs. (\ref{kapel}) and (\ref{p}).

\section{Numerical results}\label{numerical}
\begin{figure}[htb]
\psfig{figure=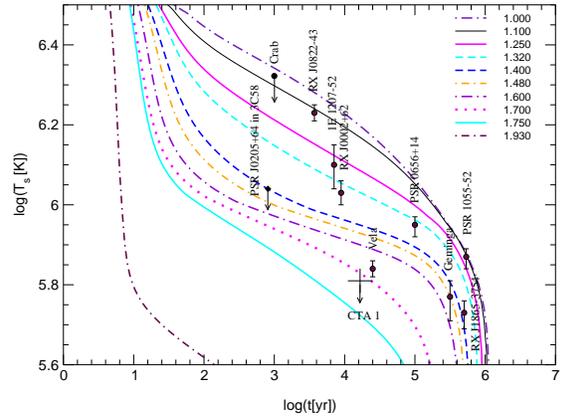,width=0.5\textwidth,angle=-90}
%%EV_HJSms.ps
\caption{Fig. 21 of \cite{BGV04}. Gaps are from Fig.  \ref{fig-gaps}
%%, thin line (Tamagaki and Takatsuka case).
for model II. The original $3P_2$ neutron
  pairing gap is  additionally suppressed by the factor $f(3P_2 ,n)=0.1$.
%%, as motivated by the result of  \cite{SF03}.
The pion gap is determined by curves 1a$+$2$+$3 of Fig. \ref{fig1}.
The $T_{\rm s} - T_{\rm in}$ relation is given by ``our fit''
curve of Fig. \ref{T-in}. Here and in all subsequent figures the
value $T_{\rm s}$ is the red-shifted temperature. NS masses are
indicated in the legend; see  \cite{BGV04}.
\label{fig21BGV}}
\end{figure}

Now we give the results of our calculations of cooling curves.
First we   present Fig. 21 of \cite{BGV04}, now Fig. \ref{fig21BGV}.
%%%%%%%%%%%%%%%%%%%%%%%%%%% Figure 15  %%%%%%%%%%%%%%%%%%
%%%%%%%%%%%%%%%%%%%%%%%%%%%%%%%%%%%%%%%%%%%%%%%%
Cooling curves shown in
 this figure were calculated using
 ``our fit'' model of the crust, shown with
the solid curve in Fig. \ref{T-in}. Here and in the corresponding
figures below the surface temperature is assumed to be
red-shifted, as  observed  at infinity from the radiation
spectrum.  Gaps are given by  model II of Fig. \ref{fig-gaps}.
However, the
%%$^3$P$_2$
$3P_2$ gap is additionally suppressed by a factor $f (3P_2
,n)=0.1$, as motivated by calculations of \cite{SF03}. 
The calculation includes pion condensation for $n>n_c^{\rm PU}$. The
figure shows a good  fit to  the data. If we used calculations
disregarding the possibility of pion condensation (see curves 1a$+$1b of Fig. 1)
we would also get an appropriate fit to the data, cf. Fig. 20 of \cite{BGV04}.
If we took
the original
%%$^3$P$_2$
$3P_2$ gap of  model II, we would not be able to describe the
data. The cooling  would be too fast, see Fig. 23 of
\cite{BGV04}. 

We will now check the  possibility of
ultra-high
%%$^3$P$_2$
$3P_2$
neutron pairing gaps, as proposed by  \cite{KCTZ04}.
In Figs. \ref{fig50,0.1} and \ref{fig50,0.5} we demonstrate the
sensitivity of the results presented in Fig. \ref{fig21BGV} to the
enhancement of the neutron $3P_2$ gap and to a suppression of the
$1S_0$ proton gap, following the suggestion of \cite{KCTZ04}. Again we use
the  calculation including pion condensation for $n>n_c^{\rm PU}$. 
\begin{figure}[ht]
%%\vspace{-0.5cm}
\centerline{
\psfig{figure=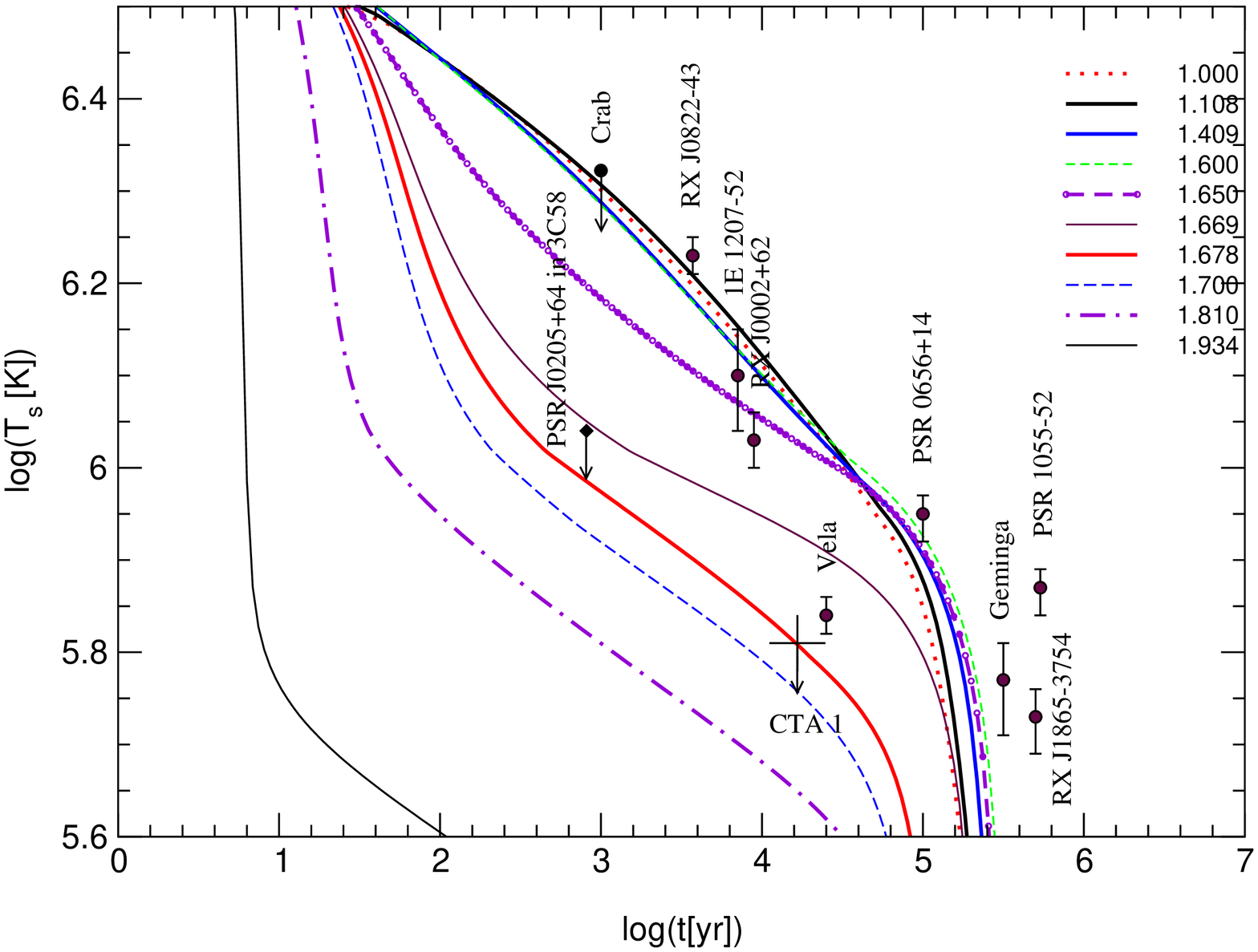,height=0.4\textwidth,angle=-90}}
\caption{Cooling curves according to the nuclear medium cooling
scenario, see Fig. \ref{fig21BGV}. Gaps are from Fig.
\ref{fig-gaps} for model II but the $3P_2$ neutron
  pairing gap is additionally enhanced  by a factor $f(3P_2 ,n)=50$ and  the
  $1S_0$ proton gap is suppressed by $f(1S_0 ,p)=0.1$.
The pion gap is determined by curves 1a$+$2$+$3 of Fig. 1.
The $T_{\rm s} - T_{\rm in}$ relation is given by ``our fit''
curve of Fig. \ref{T-in}. } \label{fig50,0.1}
\end{figure}

\begin{figure}[ht]
%%\vspace{-0.5cm}
\centerline{
\psfig{figure=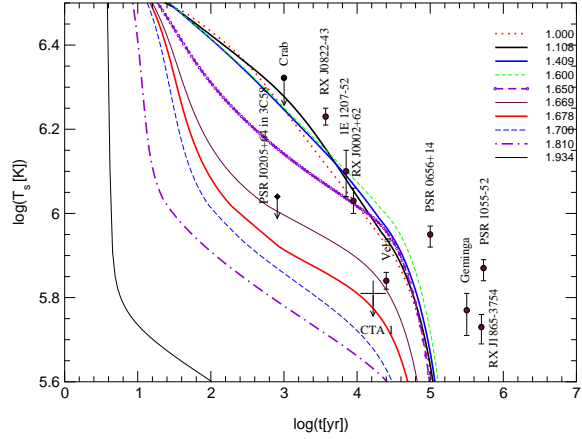,height=0.4\textwidth,angle=-90}}
\caption{Same as Fig. \ref{fig50,0.1}, but for the original $1S_0$ proton gap
suppressed by $f(1S_0 ,p)=0.5$.
} \label{fig50,0.5}
\end{figure}
We
start with   ``our crust'' model and  model II for the gaps,
using however the additional
%%We will present results of calculations with
enhancement factor $f(3P_2, n)=50$ for the neutron $3P_2$ gap.
Introducing factors $f(1S_0 ,p)=0.1$ and $f(1S_0 ,p)=0.5$ we test
the sensitivity of the results to the variation of the $1S_0$
proton gap. We do not change the value of the $1S_0$ neutron gap
since its variation almost does not influence on the cooling
curves for NSs with masses $M>1~M_{\odot}$, that we will
consider.

Comparison of Figs. \ref{fig21BGV}
%%and Figs. \ref{fig10,0.1}
 -- \ref{fig50,0.5}
shows that in all cases NSs with masses $M\gsim 1.8~M_{\odot}$
cool in similar ways in spite of the fact that $3P_2$ neutron and $1S_0$
proton gaps vary over  wide limits. This is because $3P_2$
neutron and $1S_0$ proton gaps disappear at the high densities
occuring in the central regions of these very massive NSs,
see Fig. \ref{fig-gaps}. Thus these objects cool  similarly to
non-superfluid objects. Extremely rapid cooling of stars with
$M\geq 1.84~M_{\odot}$ is due to the DU process, which is very
efficient in normal matter. The latter process appears in the central
region of NSs with $M>1.839~M_{\odot}$.
Therefore we notice that
{\em{the cooling curves are very sensitive to  the density
dependence of the gaps.}}
%%%%%From comparison of Figs. \ref{fig50,0.1} and  \ref{fig50,0.5} we see  that
The difference in the cooling of NSs with $M<1.8~M_{\odot}$ in the
cases presented in Figs. \ref{fig50,0.1} and  \ref{fig50,0.5} is
the consequence of  different values of proton gaps used in these
two calculations. This difference is mainly  due to the MpPBF
processes. The larger the proton gap, the more rapid is the
cooling.

We checked that for stars with $M\lsim 1.6~M_{\odot}$ for $T<T_{cn}$ for the
%%$^3$P$_2$
$3P_2$ neutron pairing, a complete freezing of the neutron degrees of
freedom occurs for $f (3P_2 ,n)\gsim 20$.  Then contributions
to the emissivity and to the specific heat  involving neutrons are
fully suppressed. For heavier stars ($M>1.6~M_{\odot}$) a
weak dependence on the value of the factor $f(3P_2 ,n)$ still
remains even for $f(3P_2 ,n)>100$ but the corresponding cooling
curves lie too low to allow for an appropriate fit of the data.
This difference between cooling of stars with $M<1.6~M_{\odot}$
and $M>(1.6\div 1.7)~M_{\odot}$ is due to  the mentioned density
dependence of the neutron $3P_2$ gap. The latter value smoothly
decreases with the increase of the density reaching zero for $n\gsim
4.5~n_0$ (the density $4.5~n_0$ is achieved in the center of a NS
of mass $M = 1.7~M_{\odot}$). At densities slightly below
$4.5~n_0$ the gap is  small. Therefore for stars with
$M>(1.6\div 1.7)~M_{\odot}$ the scaling of the gap by  a factor
$f(3P_2 ,n)$ changes the size of the region where  the gaps may
affect the cooling. For stars with $M\lsim 1.6~M_{\odot}$ gaps
have finite values even at the center of the star. Thus there
exists a critical value of the factor $f(3P_2 ,n)$, such that for
higher values of $f(3P_2 ,n)$ the cooling curves are 
unaffected by its change.

Figs. \ref{fig50,0.1} and  \ref{fig50,0.5} demonstrate that
we did not
succeed in reaching an appropriate overall agreement with the data; the
cooling was too rapid.
The cooling of  old pulsars is
not explained in all cases. Although the heating mechanism used by \cite{T04} may partially
help in this respect,
the discrepancy between the curves and the data points seems to be too high,
especially in Fig. \ref{fig50,0.5}.
We see that in our regime of
frozen neutron processes a better
fit is achieved in Fig.
%%%%\ref{fig10,0.1} and
\ref{fig50,0.1}, i.e.,  for a stronger suppressed proton gap (for
$f(1S_0 ,p)=0.1$). The discrepancy is even
more severe, since to justify the idea of  \cite{KCTZ04} we should
use a softer pion propagator. Only a strong softening of the
pion mode might be consistent with a significant increase of the
neutron $3P_2$ gap. On the other hand such an additional softening
would  result in  still more rapid cooling. The work
of \cite{VKZC00} discussed the possibility of a novel very efficient
process with the emissivity $\epsilon_{\nu}\propto T^5$, that
would occur due to non-Fermi liquid behavior of the Fermi sea very near
the pion condensation critical point with the
assumption of strong pion softening. If we included this very
efficient process, the disagreement with the data could be
strongly enhanced. An enhancement of the specific heat due to pion
fluctuations within the  vicinity  of the pion
condensation point cannot compensate for the acceleration of the
cooling owing to the enhancement of the emissivity. \cite{KCTZ04}
used the value $n_{c}=2~n_0$ for the critical density of the pion
condensation. In case of the Urbana-Argonne equation of state that
we exploit here (we use the HHJ fit of this equation of state that
removes the causality problem, see \cite{BGV04} for details) the
density $n=2~n_0$ is achieved in the central region of a NS with
the mass $M\sim 0.8~M_{\odot}$. This means that all NSs with $M
\gsim 0.8~M_{\odot}$ would cool extremely fast and would not be
seen in soft $X$ rays.

\begin{figure}[ht]
%%\vspace{-0.5cm}
\centerline{
\psfig{figure=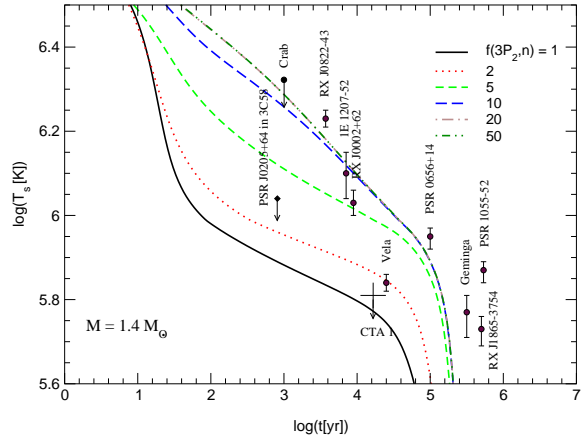,height=0.4\textwidth,angle=-90}}
\caption{Cooling curves of the neutron star with  mass
$1.4~M_{\odot}$ according to the nuclear medium cooling scenario,
see Fig. \ref{fig21BGV}. Gaps are from Fig. \ref{fig-gaps} for
model II but the $3P_2$ neutron
  pairing gap is additionally enhanced  by different factors $f(3P_2 ,n)$
  (shown in Figure)
and  the
  $1S_0$ proton gap is suppressed by $f(1S_0 ,p)=0.1$.
The pion gap is determined by curves 1a$+$2$+$3 of Fig. 1.
The $T_{\rm s} - T_{\rm in}$ relation is given by ``our fit''
curve of Fig. \ref{T-in}. } \label{figM1.4}
\end{figure}

Actually, we checked the whole interval of variation of $f(3P_2,
n)$ and $f(1S_0 ,p)$ in the range $1\div 100$ and $0.1\div
0.5$ respectively. We verified that the variation of $f(3P_2, n)$
and $f(1S_0 ,p)$  in the whole mentioned range done within
our parameterization of the effective pion gap does not 
improve the picture (curves 1a$+$2$+$3 of Fig. 1). Using  the pion gap from branches
1a$+$1b does not achieve  a better fit. In all cases we obtain too rapid cooling. To
demonstrate this, in Fig.~\ref{figM1.4} we show the cooling of a
$1.4~M_{\odot}$ star for different values of the $f(3P_2, n)$
factor. 
The factor $f(1S_0 ,p)$ is taken to be 0.1. We see that
for $f(3P_2, n)<15\div 20$ the curves rise with the increase of
$f(3P_2, n)$. For $f(3P_2, n)>20$ the curves  do not depend
on $f(3P_2, n)$.

To check how the results are sensitive to uncertainties in our
knowledge of the value  (\ref{intI}) that determines the strength
of the in-medium effect on the emissivity of the MpB process we
multiplied (\ref{MpB}) by a pre-factor $f(\rm MpB)$  that we
varied in the range $f(\rm MpB)=0.2\div 5$. In agreement with the
above discussion, for $f(\rm MpB)<1$, for temperatures $\mbox{log}
T_{\rm s}[\rm K]>5.9$ the cooling curves are shifted upwards.
For $f(\rm MpB)>1$, for temperatures $\mbox{log} T_{\rm
s}[\rm K]>5.9$ the cooling curves are shifted downwards. However,
independently of the value $f(\rm MpB)$ for $\mbox{log} T_{\rm
s}[\rm K]<5.9$ the curves are not changed. Thus it does not allow us to
diminish the discrepancy with the data.

Now we will check the efficiency of another choice of the gaps, as
motivated by model I, thick lines in Fig. \ref{fig-gaps}.
Compared to model II model I uses an {\em{artificially
enhanced proton gap}}. Thus, one can expect that model I is
less realistic than  model II. Also note  a
different density dependence of the proton gaps in model I (it cuts off at
densities $n\geq 3.2~n_0$) and in
model II (it cuts off at higher 
densities, $n\geq 4.2~n_0$).  Since uncertainties in the
existing calculations of the gaps are very high
and the 
parameterization of model I has been used by one of the groups working on the
problem of cooling of NSs, see \cite{YGKLP03}, we will  consider the
consequences of this possibility as well. 

\begin{figure}[htb]
%\vspace{-0.5cm}
\centerline{
\psfig{figure=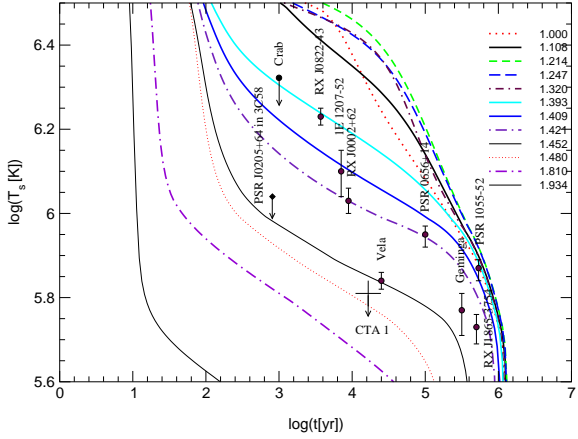,height=0.4\textwidth,angle=-90}}
\caption{Fig. 15 of \cite{BGV04}. Gaps from  model I. The
original $3P_2$  neutron gap is suppressed by $f(3P_2 ,n)=0.1$.
The pion gap is determined by curves 1a$+$2$+$3 of Fig. 1.
The $T_{\rm s} - T_{\rm in}$ relation is given by ``our fit''
curve of Fig. \ref{T-in}. For more details
  see \cite{BGV04}.} \label{fig15BGV}
\end{figure}

Fig. \ref{fig15BGV}
demonstrates our previous  fit of the data  within  model I,
but for the original $3P_2$  neutron gap  suppressed by
$f(3P_2 ,n)=0.1$, see \cite{BGV04}. Again the pion gap is determined by curves 1a$+$2$+$3 of Fig. 1.

If we took
the original
%%$^3$P$_2$
$3P_2$ gap of  model I, we would not succeed in describing the
data. The cooling  would be too fast, see Fig. 22 of
\cite{BGV04}. Therefore we   check the  possibility of
ultra-high
%%$^3$P$_2$
$3P_2$
neutron pairing gaps, as proposed by  \cite{KCTZ04}.

As  in Fig. \ref{fig50,0.1}, Fig  \ref{fig50,0.1Y}
uses $f(3P_2 ,n)=50$ and  $f(1S_0 ,p)=0.1$ but now for the gap in model
I, and, as in Fig. \ref{fig50,0.5}, Fig. \ref{fig50,0.5Y} uses $f(1S_0 ,p)=0.5$
for the gap in model I.
Figs. \ref{fig50,0.1Y} and \ref{fig50,0.5Y} show that within the
variation of the gaps of  model I the discrepancy  to the data is still
stronger compared to that for the above calculation  based on the use of  model II.
The difference between curves shown in Figs. \ref{fig50,0.1Y} and
\ref{fig50,0.5Y} is less pronounced than for those curves
in Figs.  \ref{fig50,0.1} and  \ref{fig50,0.5}.
Indeed,  as we have mentioned, the density dependence of the
proton gap is different in models I and II. In model II  the
proton gap reaches  higher densities ($\simeq 4.2~n_0$)
than in  model I ($\simeq 3.2~n_0$). Thus
in the  case shown by Figs. \ref{fig50,0.1Y} and  \ref{fig50,0.5Y}
a non-superfluid core begins to contribute at lower
values of the star mass.
%%for stars with $M\gsim 1.3~M_{\odot}$  essentially contributes to the cooling
%%Figs. \ref{fig50,0.1Y} and  \ref{fig50,0.5Y} demonstrate
Therefore  in both figures the corresponding cooling curves are almost the same
for  $M\gsim 1.6 M_{\odot}$. Using the pion gap from branches
1a$+$1b, i.e. disregarding the possibility of pion condensation, does not
allow a better fit.

\begin{figure}[ht]
%%\vspace{-0.5cm}
\centerline{
\psfig{figure=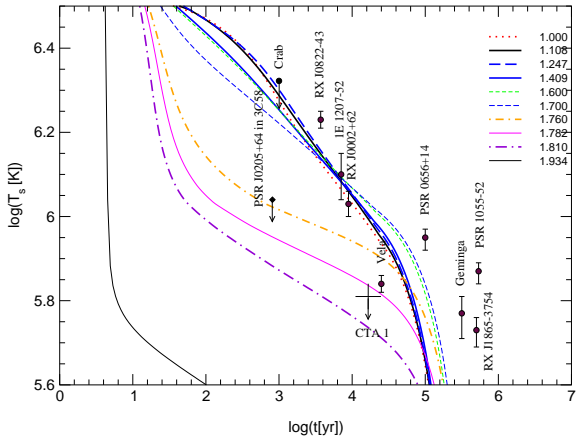,height=0.4\textwidth,angle=-90}}
\caption{Cooling curves according to the nuclear medium cooling
scenario, see Fig. \ref{fig15BGV}. Gaps are from Fig.
\ref{fig-gaps} for model I but the $3P_2$ neutron
  pairing gap is additionally enhanced  by a factor $f(3P_2 ,n)=50$ and  the
  $1S_0$ proton gap is suppressed by $f(1S_0 ,p)=0.1$.
The pion gap is determined by curves 1a$+$2$+$3 of Fig. 1.
The $T_{\rm s} - T_{\rm in}$ relation is given by ``our fit''
curve of Fig. \ref{T-in}. } \label{fig50,0.1Y}
\end{figure}
\begin{figure}[ht]
%%\vspace{-0.5cm}
\centerline{
\psfig{figure=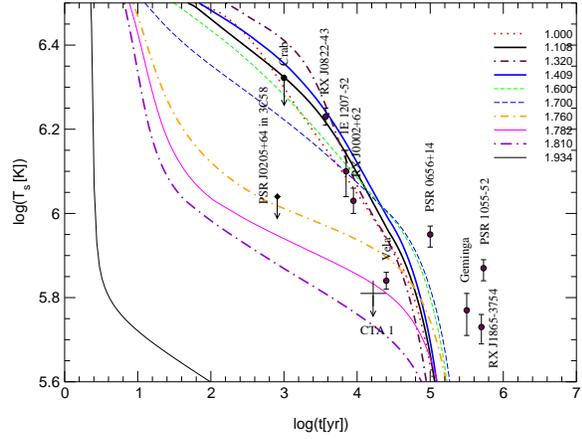,height=0.4\textwidth,angle=-90}}
\caption{Same as Fig. \ref{fig50,0.1Y}, but for the original
 $1S_0$ proton gap suppressed by $f(1S_0 ,p)=0.5$.
} \label{fig50,0.5Y}
\end{figure}

The dependence of the results on the different choices of the
$T_{\rm s} - T_{\rm in}$ relation is demonstrated  by Figs.
\ref{fig10,0.1etII} and \ref{fig50,0.1cr}
%%, \ref{fig10,0.5etII}
for gaps based on a modification of model II.
For this demonstration we first took the upper boundary curve
$\eta =4\cdot 10^{-16}$ and then the lower boundary curve $\eta
  =4.0 \cdot 10^{-8}$ in Fig.~\ref{T-in}.
We show that these choices however do not allow us to improve the
fit. Comparing Figs. \ref{fig10,0.1etII} ($\eta =4\cdot 10^{-16}$) and
\ref{fig50,0.1} (``our fit'')
based on the  same modification of  model II we see that
with the ``our fit'' crust model (Fig. \ref{fig50,0.1}) the deviation from the data
points is less pronounced. We have checked that based on 
model I
one arrives at the same conclusion. An increase of the factor $f(1S_0
,p)$ to $0.5$ reduces the  fit.
\begin{figure}[ht]
\vspace{-0.2cm} \centerline{
\psfig{figure=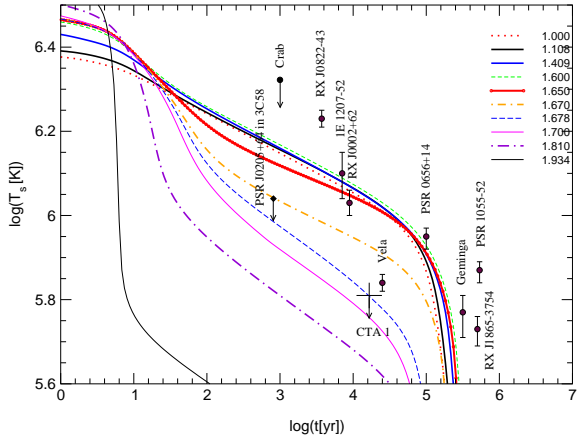,height=0.4\textwidth,angle=-90}}
\caption{Cooling curves according to the nuclear medium cooling
scenario. Gaps are from Fig.
\ref{fig-gaps} for model II but the $3P_2$ neutron
  pairing gap is additionally enhanced  by a factor $f(3P_2 ,n)=50$ and  the
  $1S_0$ proton gap is suppressed by $f(1S_0 ,p)=0.1$.
The pion gap is determined by curves 1a$+$2$+$3 of Fig. 1.
The $T_{\rm s} - T_{\rm in}$ relation is given by  the crust
model for $\eta
  =4.0 \cdot 10^{-16}$.
} \label{fig10,0.1etII}
\end{figure}
\begin{figure}[ht]
%%\vspace{-0.5cm}
\centerline{
\psfig{figure=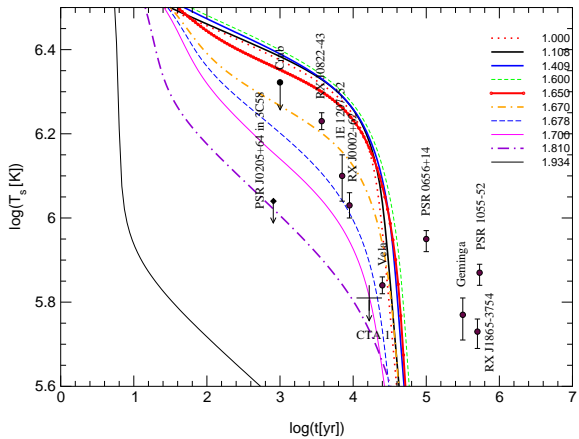,height=0.4\textwidth,angle=-90}}
\caption{Same as in Fig. \ref{fig10,0.1etII},
%\ref{fig50,0.1}
 but for the crust
model $\eta
  =4.0 \cdot 10^{-8}$.
} \label{fig50,0.1cr}
\end{figure}
In Fig. \ref{fig50,0.1cr} we use the lower boundary curve $\eta
  =4.0 \cdot 10^{-8}$ of  Fig. \ref{T-in}.
We further demonstrate that the selection of a different choice of
the $T_{\rm s}-T_{\rm in}$ relation within the band shown in
Fig.~\ref{T-in} does not diminish the discrepancy. This
  discrepancy increases compared to that demonstrated by   ``our fit''
  model.  Indeed, the cooling evolution for times $t\lsim 10^5$~yr
($T_{\rm s}\gsim
  10^{6}$K) is governed by neutrino processes. Thus the higher $T_{\rm in}$,
the larger $T_{\rm s}$ is. The slowest cooling is then  obtained, if one
  uses the lower boundary curve $\eta
  =4.0 \cdot 10^{-8}$ of Fig.~\ref{T-in}.
The evolution of NSs for   times $t\gsim 10^5$~yr begins to be controlled by
  the photon processes. In the photon epoch ($t\gg 10^5$~yr) the smaller the  $T_{\rm s}$ value, the
  less efficient the radiation is. Thus for $t\gg 10^5$~yr
the slowest cooling is obtained if one uses the upper boundary curve $\eta
  =4.0 \cdot 10^{-16}$ of Fig.~\ref{T-in}. The ``our crust'' curve  simulates the
  transition from  one limiting curve to the other
demonstrating the slowest cooling in the  temperature
interval shown in the figures.

{\em In all cases the data are not explained by
the assumption of an enhanced neutron $3P_2$ gap (for $f(3P_2 ,n)
>1$) and a partially suppressed $1S_0$ proton gap (for $f(1S_0
,p)=0.1\div 0.5$).}

\section{Concluding remarks}\label{conclusion}

Our aim was to consider  large $3P_2$ gaps
within the "nuclear medium cooling" scenario of \cite{BGV04}
that well described the cooling data with the opposite assumption of
suppressed $3P_2$ gaps. Therefore  we did not
incorporate  extra assumptions of internal heating for old pulsars, see
\cite{T04}, and  quark cores in massive NSs, see
\cite{BGVquark1} and \cite{BGVquark2} and refs therein.

The main problem with the given scenario is that at
the frozen neutron contribution to the specific heat and to the
emissivity, the region of surface temperatures $T_{\rm s} > 10^6$K
is determined by proton processes. The most efficient among them
is the MpPBF process for $T<T_{cp}$ and MpB for $T>T_{cp}$. For the
proton gaps that we deal with, the
MpPBF process proves to be too efficient, yielding too rapid
cooling. Thus at least several slow cooling data points (at least
in data for old pulsars) are not explained. Some works
ignore the medium-induced enhancement of the MpPBF
emissivity that results in a 10-100 times suppression of the rate.
We omitted this possibility as physically unrealistic. The origin of this
enhancement is associated with opening up of new reaction channels
in the medium that are forbidden in vacuum.

Thus, we have shown that the ``nuclear medium cooling''
scenario of \cite{BGV04} fails to appropriately fit the neutron star cooling
data with the assumption of a strong enhancement of the $3P_2$
neutron gaps (we checked the range  $f(3P_2 ,n)=1\div 100$) and
for moderately suppressed $1S_0$ proton gaps (for $f(1S_0
,p)=0.1\div 0.5$). On the other hand the  same scenario
allowed us to appropriately fit the whole set of data  with the
assumption of a significantly suppressed $3P_2$ neutron gap (for
$f(3P_2 ,n) \sim 0.1$). We observed  essential dependence of the
results not only on the values of the gaps but also on their
density dependence. We used the density dependence of the gaps
according to  models I and II. The latter model  is supported
by microscopic calculations. We excluded an attempt to
artificially fit  the density dependence of the gaps trying to
match cooling curves with the data. Although such an attempt could
improve the fit, it is not physical and we did
not pursue it. However we  encourage further
attempts of microscopic  calculations of the gaps, which would
take into account the most important medium effects. With correctly
treated gaps one could perform new simulations of  NS cooling.

{\it{Acknowledgements.}}
We thank D. Blaschke for his  interest in our work,
critical reading of the manuscript and valuable remarks. We also
thank B. Friman and A. Schwenk for interesting discussions and
A.Y. Potechin for helpful comments. The research of H.G. was
supported by the Virtual Institute of the Helmholtz Association
under grant No. VH-VI-041 and by the DAAD partnership program
between the Universities of Rostock and Yerevan. The work of
D.N.V. was supported in part by the Deutsche
Forschungsgemeinschaft (DFG project 436 RUS 113/558/0-2) and the
Russian Foundation for Basic Research (RFBR grant 03-02-04008).

\end{fmffile}
%{medneutrd}
\end{document}